**Transition from Homogeneous to Domain-Wall-Mediated Polarization Switching in $BaTiO_3$: A Machine-Learning Molecular Dynamics Study**

Po-Yen Chen,[1,*] Teruyasu Mizoguchi,[1,2,†]

[1]Department of Materials Engineering, The University of Tokyo, 153-8505 Tokyo, Japan.
[2]Institute of Industrial Science, The University of Tokyo, 153-8505 Tokyo, Japan.

**ABSTRACT**. Polarization switching in ferroelectric $BaTiO_3$ can proceed through fundamentally different mechanisms — yet the conditions that determine which pathway is realized remain poorly understood. Using machine-learning potential-based molecular dynamics with the MACEField model, we systematically vary supercell size to reveal a clear transition from homogeneous polarization switching to domain-wall (DW)-mediated switching, accompanied by a coercive field increase of over 50%. Shannon entropy analysis demonstrates that this transition is driven by size-dependent polarization fluctuations that promote 180° DW nucleation — establishing a direct, quantitative link between local configurational disorder and macroscopic switching behavior. Additional analyses further suggest that the switching pathway is also influenced by the kinetic competition between DW nucleation/growth and annihilation. Furthermore, the switching pathway and hysteresis response are shown to depend critically on supercell geometry and the relative orientation of applied stress and electric field. These findings reveal that homogeneous and DW-mediated switching are distinct physical regimes in $BaTiO_3$, and that atomistic simulations must account for system size to correctly capture the operative switching mechanism.

## I. INTRODUCTION.

Ferroelectric polarization behavior plays a crucial role in determining the functional properties of ferroelectric materials, including dielectric response, piezoelectricity, and non-volatile switching characteristics[1-5]. Among various ferroelectric materials, tetragonal $BaTiO_3$ has been extensively studied as a prototypical perovskite ferroelectric owing to its well-known spontaneous polarization, multiple structural phase transitions, and strong electromechanical coupling[6-10]. In tetragonal $BaTiO_3$, ferroelectricity originates from the off-center displacement of Ti ions within the oxygen octahedra, leading to spontaneous polarization along the crystallographic c-axis. The polarization state is strongly coupled to lattice distortion and domain configuration, making polarization switching highly sensitive to external electric fields and mechanical constraints[11-14].

Despite extensive experimental [15-20] and computational investigations[11, 21-23], a fundamental question remains open: polarization switching in $BaTiO_3$ can proceed through qualitatively different pathways, homogeneous reversal or domain-wall (DW)-mediated switching, yet the conditions that determine which mechanism is operative are poorly understood. In atomistic simulations, homogeneous switching refers to a collective polarization reversal throughout the simulation cell, accompanied by coherent lattice deformation and frequently proceeding through a two-step polarization-rotation pathway with intermediate polarization states[11, 21]. However, in finite periodic supercells, this switching behavior may also arise because DW nucleation and propagation are suppressed by the limited simulation-cell size and periodic boundary conditions, leading to the nearly homogenous switching. In contrast, DW-mediated switching proceeds through local domain nucleation and subsequent DW propagation[12, 22-24]. These two mechanisms represent fundamentally different microscopic and macroscopic polarization dynamics and therefore lead to distinct coercive fields, switching kinetics, and electromechanical responses. Therefore, clarifying the competition between homogeneous and DW-mediated switching is essential for understanding the intrinsic ferroelectric behavior of $BaTiO_3$ and for improving the performance of ferroelectric-based devices.

Previous atomistic simulations have reported apparently different switching behaviors depending on the simulation conditions: NVT-based simulations[22] tend to favor domain nucleation and DW formation, whereas NPT-based simulations[11, 21] more frequently exhibit homogeneous polarization switching. However, it remains unclear whether these differences primarily originate from the choice of simulation ensemble or reflect an intrinsic physical crossover in ferroelectric switching behavior. Clarifying this distinction is essential not only for interpreting atomistic simulations, but also for understanding the microscopic origin of coercive fields and switching kinetics in real ferroelectric systems.

The stability of homogeneous and DW-mediated switching is expected to strongly depend on the accessible length scale, strain accommodation, and geometric constraints of the system. In our previous study,[11] based on NPT simulations with fully relaxable lattice degrees of freedom, we observed that larger $BaTiO_3$ supercells under compressive stress conditions tend to stabilize 180° DW structures during polarization switching, suggesting a crossover from homogeneous collective switching to domain-mediated polarization reversal as the system size increases. Nevertheless, the relationship between supercell size, domain stability, and switching mechanism has not yet been systematically clarified.

*Contact author: poyen@iis.u-tokyo.ac.jp
†Contact author: teru@iis.u-tokyo.ac.jp

Therefore, in the present study, we systematically investigate the influence of supercell size and geometry on polarization switching behavior in tetragonal $BaTiO_3$ using machine learning potential-based molecular dynamics (MLP-MD) simulations under external electric fields within the NPT ensemble. By varying the accessible length scale under strain-relaxing conditions, we analyze hysteresis loops and polarization switching processes. Particular attention is given to the emergence of DW-mediated switching, its onset conditions, and its influence on ferroelectric properties, including coercive fields, switching pathways, and electromechanical responses in $BaTiO_3$.

## II. METHODS

### A. MLP and BEC model

In this study, we employed the MLP-MD framework developed in our previous works[11, 25]. The interatomic interactions were described using our previously finetuned MACE model[26], while the Born effective charge (BEC) was predicted using the MACEfield-finetuned model[27]. Compared with the previously developed Equivar model[28, 29], the MACEfield model provides significantly faster BEC calculations for large supercell systems. The computational performance of the MACEfield model is summarized in Appendix A. The finetuned MACE model was developed based on the pretrained MACE-MP-0 foundation model[30, 31] and further finetuned using our $BaTiO_3$ dataset containing 4,045 2×2×2 $BaTiO_3$ structures with the PBEsol functional[26]. Since the present study focuses on solid-state $BaTiO_3$ systems, a subset containing 2,038 solid-phase structures with BEC data[11] calculated by density functional perturbation theory (DFPT)[32, 33] implemented in Vienna Ab initio Simulation Package (VASP)[34-36] was selected from the database for the MACEfield training. The MACEfield model was then further finetuned based on the previously trained MACE model. The final MACEfield model achieved a mean absolute error (MAE) of approximately 0.01 for the BEC prediction of solid-state structures, which is significantly smaller than the corresponding MAE of 0.03 obtained using the previous Equivar model.

### B. MLP-MD workflow and hysteresis-loop simulations

After constructing the MLP and BEC models, we employed the MLP-MD framework to investigate the electric-field response of ferroelectric tetragonal $BaTiO_3$ systems. During each MD step, the total force acting on atoms was calculated as $F_{total} = F_{MLP} + Z \cdot \varepsilon$, where $F_{MLP}$ is the interatomic force predicted by the MLP, $Z$ is the BEC, and $\varepsilon$ is the external electric field. This framework was implemented within the atomic simulation environment (ASE)[37] package under the NPT ensemble to enable MD simulations under finite electric fields. To further accelerate the MD simulations and improve computational efficiency for large supercell systems, cuEquivariance acceleration[38] was employed during the MACE/MACEField calculations. The computational performance improvement obtained using cuEquivariance is discussed in Appendix A.

For the P–E hysteresis-loop simulations, we employed the Nosé–Hoover thermostat together with the Parrinello–Rahman barostat within the NPT ensemble[39, 40]. In the present study, the maximum electric-field amplitude was increased to 150 kV/cm, compared with 100 kV/cm in our previous work[11]. To maintain a comparable electric-field sweeping rate (1 kV/(cm*ps)) on the hysteresis behavior, the total MD simulation steps were increased from 400,000 to 600,000 steps, corresponding the frequency of 1.67 GHz in the simulation. The sweep rate dependent hysteresis behavior investigation is conducted in Appendix B.

### C. Ti-displacement calculations

In order to analyze the local polarization behavior in $BaTiO_3$, we evaluated the displacement of Ti atoms relative to the centers of their surrounding oxygen coordination environments. In tetragonal $BaTiO_3$, spontaneous polarization originates primarily from the off-center displacement of Ti atoms within TiO6 octahedra. Therefore, the local polarization direction can be characterized using the Ti off-center displacement vector. For each Ti atom, neighboring oxygen atoms within a cutoff radius of 3 Å were identified using a neighbor-search algorithm. The geometric center of the surrounding oxygen atoms was calculated as $\mathbf{R}_{env,O} = \frac{1}{N}\sum_{i=1}^{N}\mathbf{R}_{i,O}$, where $N$ is the number of neighboring oxygen atoms and $\mathbf{R}_{i,O}$ denotes the Cartesian coordinates of each neighboring oxygen atom. The local Ti-displacement vector was then defined as $\Delta\mathbf{R}_{Ti} = \mathbf{R}_{Ti} - \mathbf{R}_{env,O}$, where $\mathbf{R}_{\mathrm{Ti}}$ is the Ti atomic position. This displacement vector was used to characterize the local polarization direction

*Contact author: poyen@iis.u-tokyo.ac.jp
†Contact author: teru@iis.u-tokyo.ac.jp

and analyze polarization distributions and domain structures in $BaTiO_3$.

## III. RESULTS and DISCUSSION

### A. Size-dependent P-E hysteresis loop and micro polarization behavior along the b-axis electric field

First, we investigated the evolution of polarization switching behavior with increasing supercell size along the c axis. Since the stress-induced DW structures preferentially formed along the a/b-axis orientation in our previous observations[11], and considering the symmetry of the tetragonal structure, triangular electric fields were first applied along the b axis. The maximum electric field amplitude and sweep rate were set to 150 kV/cm and 1 kV/(cm*ps), respectively. To examine the finite-size effect, the supercell size was systematically increased from 8×8×8 to 8×8×32. The resulting P–E hysteresis loops and coercive fields are presented in Figure 1a and b. Although the remanent polarization remains nearly unchanged with increasing supercell size, a significant increase in the averaged coercive field is observed for supercells larger than 8×8×16. Specifically, the averaged coercive fields obtained from three independent simulations are 62.9 ± 9.2, 78.4 ± 9.2, 81.8 ± 3.2, and 86.8 ± 6.0 kV/cm for the 8×8×8, 8×8×16, 8×8×24, and 8×8×32 supercells, respectively. The results reveal a clear increase in the coercive field with increasing supercell size, with the smallest 8×8×8 supercell exhibiting a significantly lower coercive field compared with the larger supercells. Beyond the 8×8×16 supercell, the coercive field shows a relatively weak size dependence, suggesting that the switching behavior approaches a converged regime for larger system sizes. This distinct change in coercive behavior suggests that the underlying polarization switching mechanism in larger supercells differs from that in the smaller 8×8×8 system.

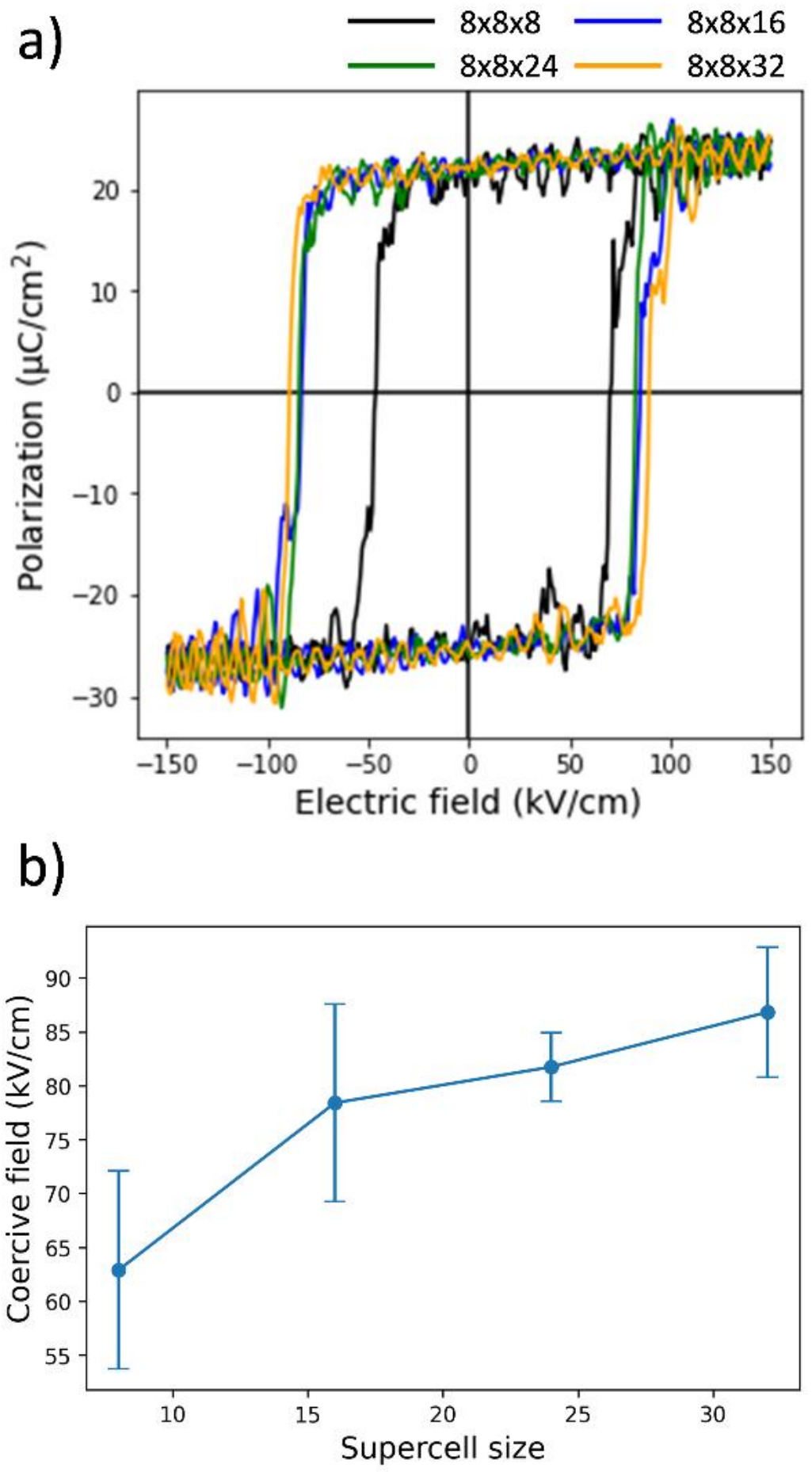

**Figure 1** Size dependence of (a) hysteresis loops and (b) coercive field for 8 × 8 × n tetragonal $BaTiO_3$ supercells at 250 K under zero mechanical stress. The black, blue, green, and orange curves correspond to n = 8, 16, 24, and 32, respectively. Error bars indicate the standard deviation of the coercive field obtained from independent simulations.

To further investigate the polarization-switching mechanism, we analyzed the local polarization based on Ti displacements. Figure 2 presents the local polarization projected onto the bc plane in the 8×8×32 supercell during polarization switching, where the arrows indicate the local polarization directions. As shown in the figure, local polarization reversal first emerges inside the supercell (step 397k), indicating the nucleation of an oppositely polarized domain. With continued electric-field application, the nucleated domain gradually expands (steps 398k and 399k) and eventually produces complete polarization reversal (step 450k) through the formation and propagation of a 180° DW.

*Contact author: poyen@iis.u-tokyo.ac.jp
†Contact author: teru@iis.u-tokyo.ac.jp

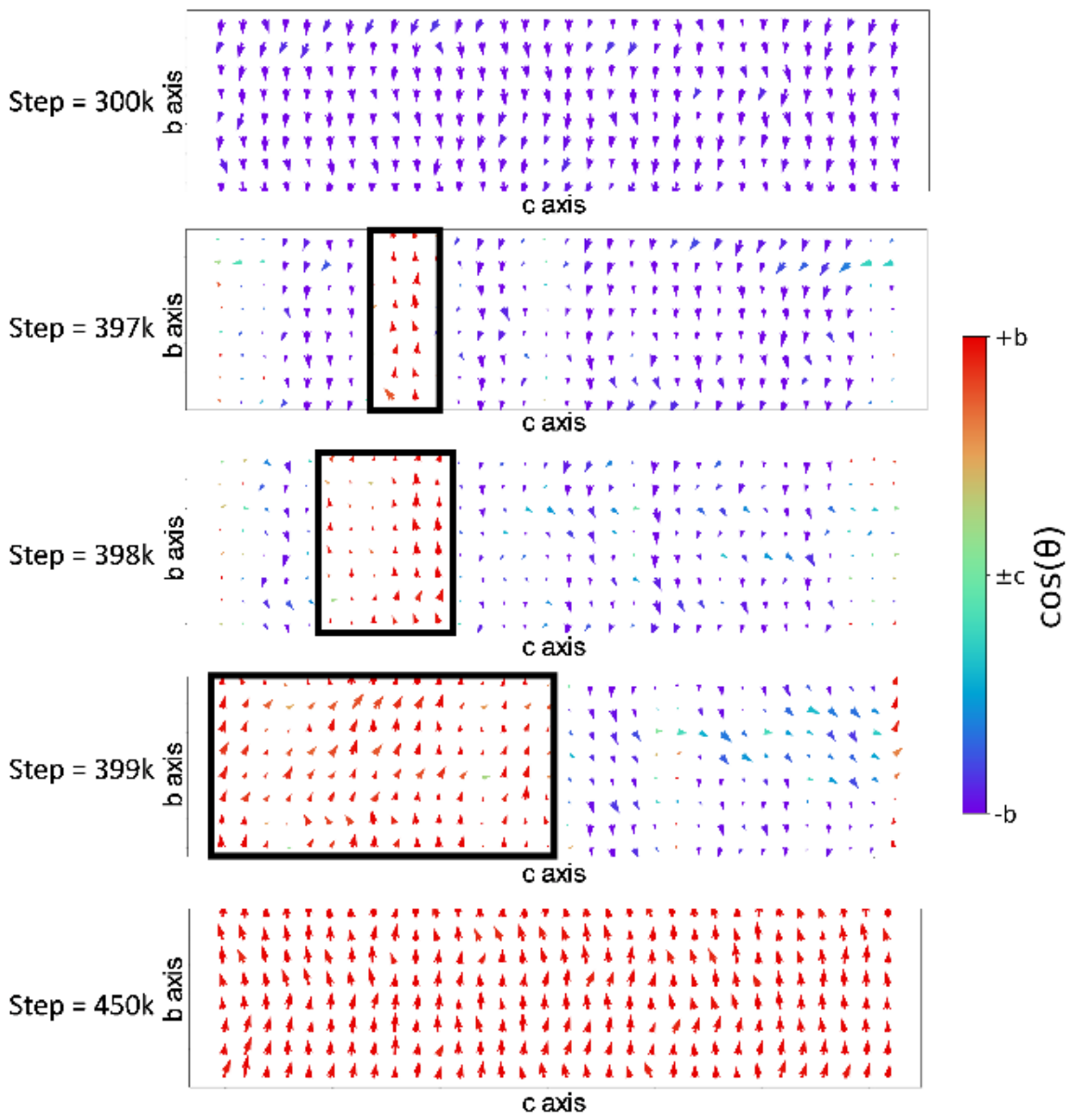

**Figure 2** Evolution of Ti displacement vectors in the 8×8×32 supercell at selected MD steps under a triangular electric field (maximum field: 150 kV/cm) along the *b* axis at 250 K. The arrows show the Ti-displacements on each Ti atom, and the boxed region shows the opposite domain nuclei.

As discussed in our previous study on stress-induced DW formation[11], larger supercells exhibit a higher probability of forming DW structures during polarization switching. To further quantify this behavior, we employed Shannon entropy analysis[41] to characterize the degree of polarization fluctuation when polarization switching occurs. The calculation for Shannon entropy is mentioned in Appendix C. As shown in Figure 3, the maximum Shannon entropy during simulation increases with supercell size, indicating enhanced polarization fluctuations in larger systems when the polarization switching occurs. Such increased fluctuations may facilitate the local instability required for DW nucleation. Consequently, larger supercells tend to favor DW-mediated switching pathways, leading to polarization reversal through domain nucleation and growth. In contrast, smaller supercells exhibit reduced polarization fluctuations, which suppress local nucleation events and instead favor homogeneous polarization switching.

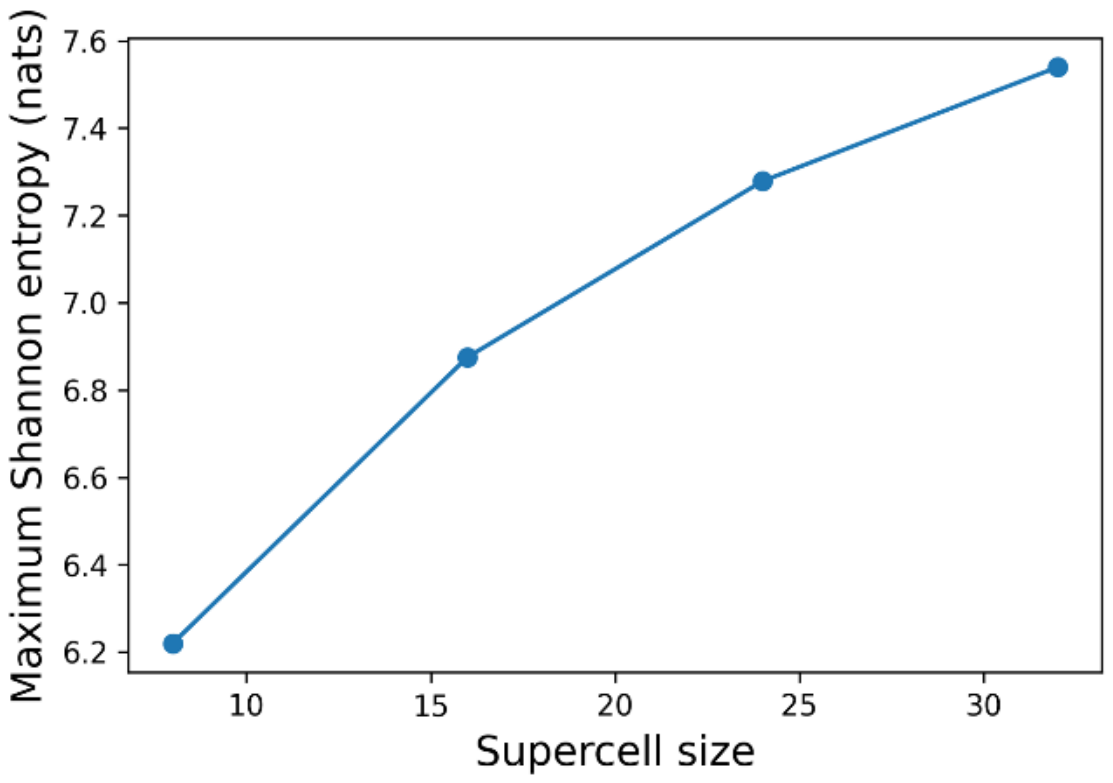

**Figure 3** Size dependent maximum Shannon Entropy during simulation of the 8×8×n tetragonal $BaTiO_3$ supercell at 250 K.

Whereas, because the elongated supercells were constructed by extending only the c-axis, the effects of increasing system size and changing the simulation-cell aspect ratio cannot be completely separated. To clarify whether the observed change in the switching mechanism originates from the system size or the aspect ratio, we further investigated the coercive field using isotropic supercells with dimensions ranging from 8 × 8 × 8 to 12 × 12 × 12, as shown in Figure 4a. The coercive field increases with increasing isotropic supercell size and approaches a converged value of approximately 100 kV/cm for supercells of 10 × 10 × 10 and larger. In addition, the local polarization evolution of the 10 × 10 × 10 supercell, shown in Figure 4b, reveals the emergence of DW-mediated switching. These results indicate that the qualitative transition from homogeneous switching to DW-mediated switching is primarily governed by increasing system size rather than by the elongated simulation-cell geometry.

*Contact author: poyen@iis.u-tokyo.ac.jp
†Contact author: teru@iis.u-tokyo.ac.jp

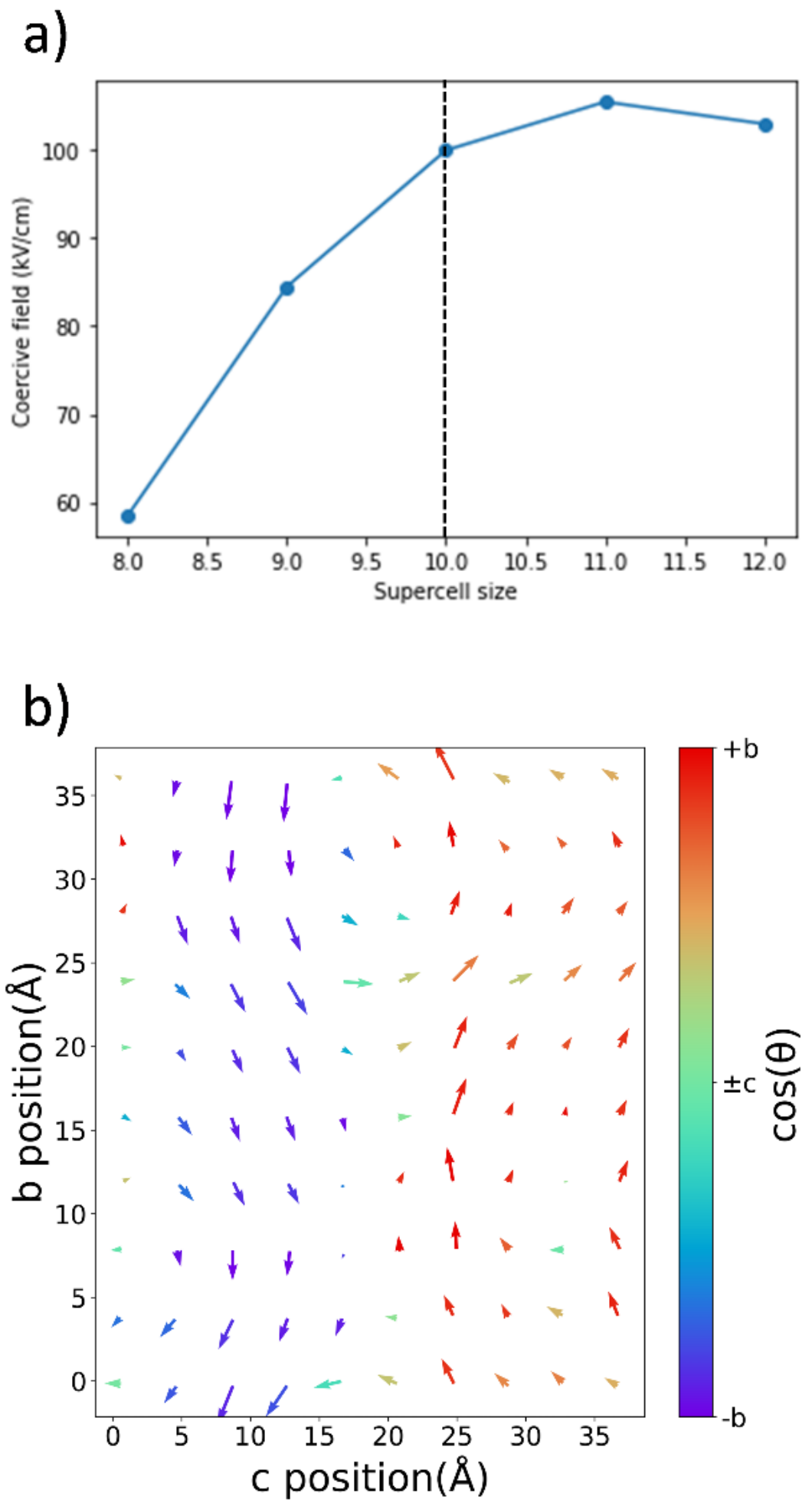

**Figure 4** a) Size dependence of the coercive field for n × n × n isotropic supercells. The dashed line indicates the smallest supercell that exhibits DW-mediated switching. b) Ti displacement vectors in the 10 × 10 × 10 supercell at the 400,000th MD step during the simulation at 250 K.

The size dependence of the coercive field further suggests two distinct switching regimes. For relatively small supercells, homogeneous switching requires the cooperative rotation of all dipoles, and the associated switching barrier therefore increases with system size. Once the supercell becomes sufficiently large for DW nucleation to occur, the switching process is instead governed by the nucleation barrier, which can be expressed as $E_{nucleation} = \frac{c\sigma_{dw}^2}{P_{sat}k_BT}$,[42] where $c$ is the DW width, $\sigma_{dw}$ is the DW energy, $P_{sat}$ is the saturated polarization. Since these quantities are intrinsic material properties and are essentially independent of the simulation-cell size, the coercive field becomes nearly converged once the supercell is sufficiently large for DW nucleation to become energetically favorable. This observation is also consistent with the results for the anisotropic supercells shown in Figure 1b, indicating that the emergence of DW-mediated switching and the convergence of the coercive field are governed by the supercell size rather than by the large aspect ratio.

Although the DW-mediated switching observed here is qualitatively similar to that reported by Falletta et al.[22], the underlying switching pathway is fundamentally different. Falletta et al. performed their simulations under the NVT Langevin simulations, in which the simulation-cell shape is fixed and cannot deform during polarization switching. To directly isolate the influence of the statistical ensemble, we additionally performed Langevin NVT simulations (friction coefficient = 0.01) using the same 8 × 8 × 8 supercell, MACEField potential, temperature, and electric-field protocol employed in the present NPT simulations as shown in Figure 5a. Under the same electric-field conditions, no polarization switching was observed in the NVT ensemble. Polarization switching occurred only after increasing the electric-field amplitude by a factor of two, corresponding to an average coercive field of approximately 160 kV/cm.

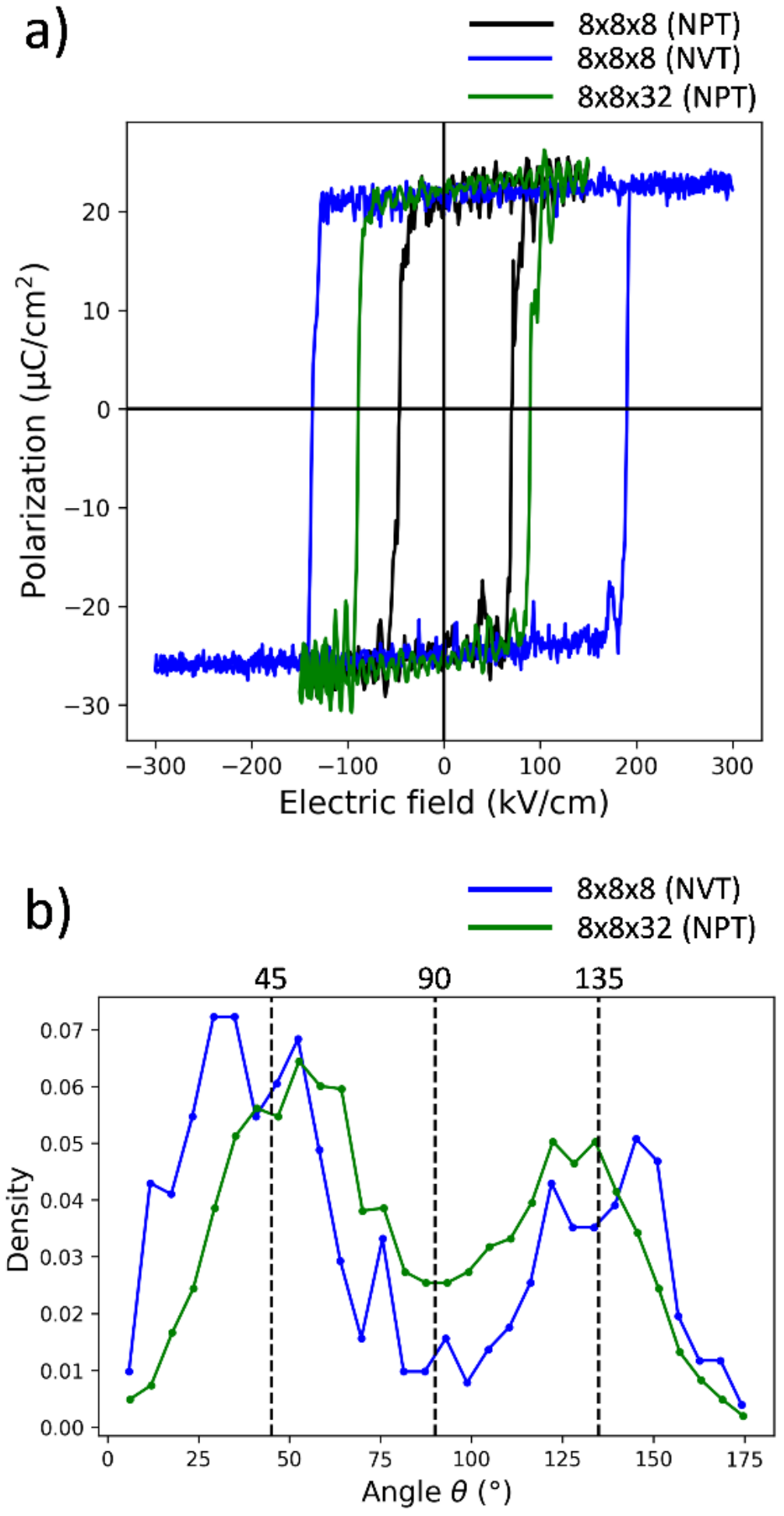

*Contact author: poyen@iis.u-tokyo.ac.jp
†Contact author: teru@iis.u-tokyo.ac.jp

**Figure 5** a) Hysteresis loops of the 8 × 8 × 8 supercell obtained using the NPT (black) and NVT (blue) ensembles, together with the hysteresis loop of the 8 × 8 × 32 supercell obtained using the NPT ensemble (green). b) Distribution of Ti displacement angles with respect to the polarization direction.

Further insight is obtained from the polarization-angle distribution during switching as shown in Figure 5b. The NVT simulations exhibit a negligible population around 90°, indicating that polarization reversal proceeds predominantly through a direct-reversal pathway, consistent with the observations of Falletta *et al.*[22]. In contrast, the NPT simulations with an 8 × 8 × 32 supercell exhibit an amount of population around 90°, demonstrating that polarization rotation remains an essential component of the switching process even when DW-mediated switching is present. These results indicate that DW formation and polarization rotation are not mutually exclusive under the NPT ensemble. Instead, the large supercells exhibit DW-mediated switching via the rotation pathway, whereas the NVT simulations favor DW-mediated switching via direct polarization reversal because the constrained simulation-cell shape suppresses lattice deformation.

This interpretation is further supported by previous NPT-based molecular dynamics simulations[11, 21], Landau–Ginzburg–Devonshire calculations[24], and experimental observations[17-19] , all of which suggest that polarization switching preferentially proceeds through the rotation pathway because it possesses a lower activation barrier than direct polarization reversal[24]. Since polarization rotation is intrinsically coupled to lattice deformation, the mechanical constraints imposed by the NVT ensemble suppress this low-energy pathway, leading to a substantially larger coercive field than that obtained under the NPT ensemble.

**B. Temperature-dependent P-E hysteresis loop and polarization switching mechanism**

Because the previous section suggests that DW-mediated switching is associated with the Shannon entropy, which is expected to be influenced by thermal fluctuations, we further investigated the temperature dependence of the switching behavior. To this end, simulations were performed using an 8×8×24 supercell at temperatures ranging from 230 to 290 K. The simulated hysteresis loops and the corresponding maximum Shannon entropy during the switching process are shown in Figure 6a and b, respectively. As shown in the figure, the coercive field decreases with increasing temperature because enhanced thermal fluctuations facilitate polarization reversal. Meanwhile, the maximum Shannon entropy also increases with temperature, reflecting the enhanced polarization fluctuations.

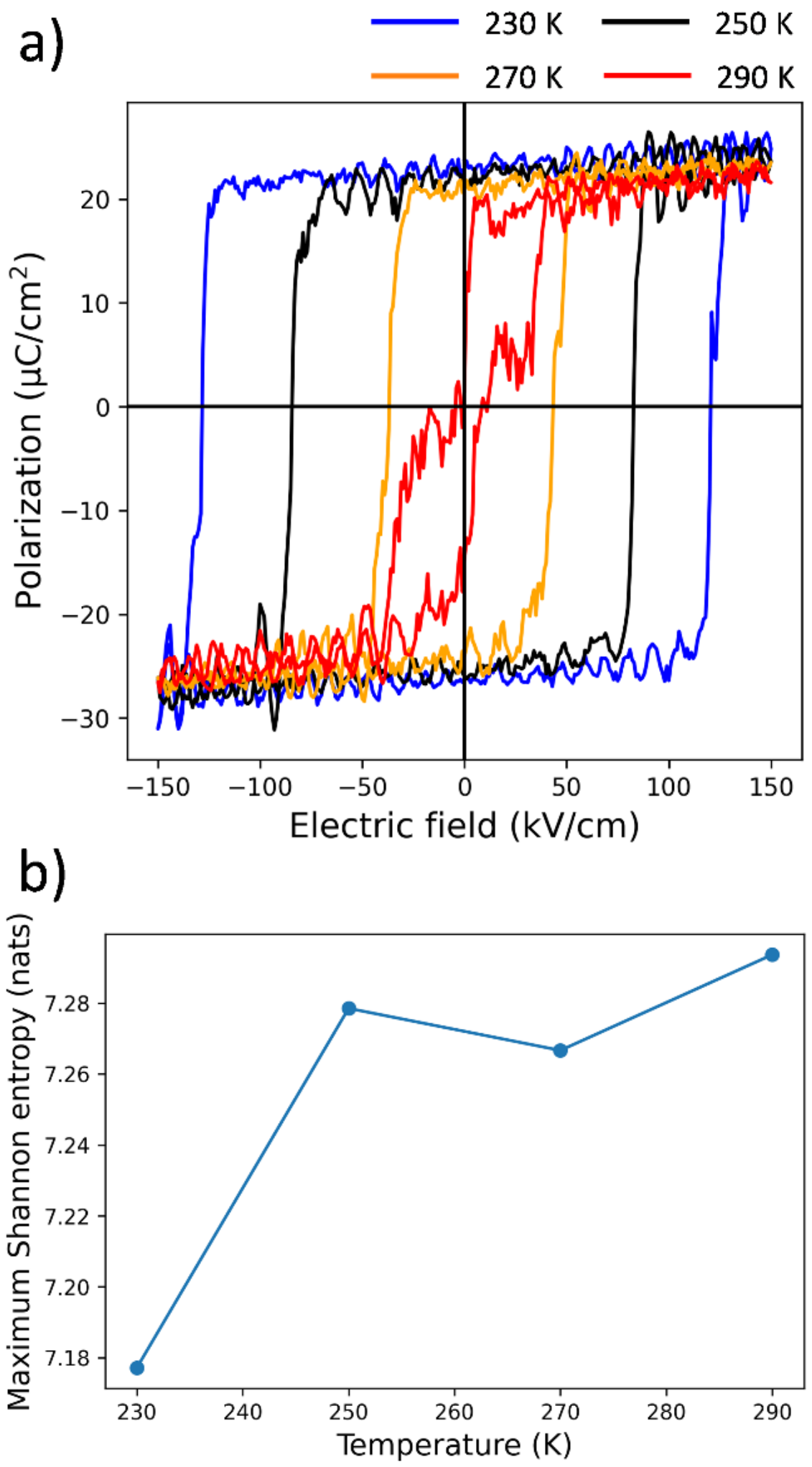

**Figure 6** Temperature dependent a) hysteresis loops and b) maximum Shannon entropy of the 8 × 8 × 24 supercell.

However, when examining the switching pathway at 270 K, the system still exhibits a nearly homogeneous rotation pathway rather than DW-mediated switching, as shown in Figure 7. During the initial stage of polarization reversal, a small region with reversed polarization is indeed observed. Nevertheless, this transient reversed polarization region rapidly disappears due to thermal fluctuations, resulting in a nearly homogeneous switching process. This observation indicates that a high Shannon entropy alone is not a sufficient condition for triggering DW-mediated switching.

*Contact author: poyen@iis.u-tokyo.ac.jp
†Contact author: teru@iis.u-tokyo.ac.jp

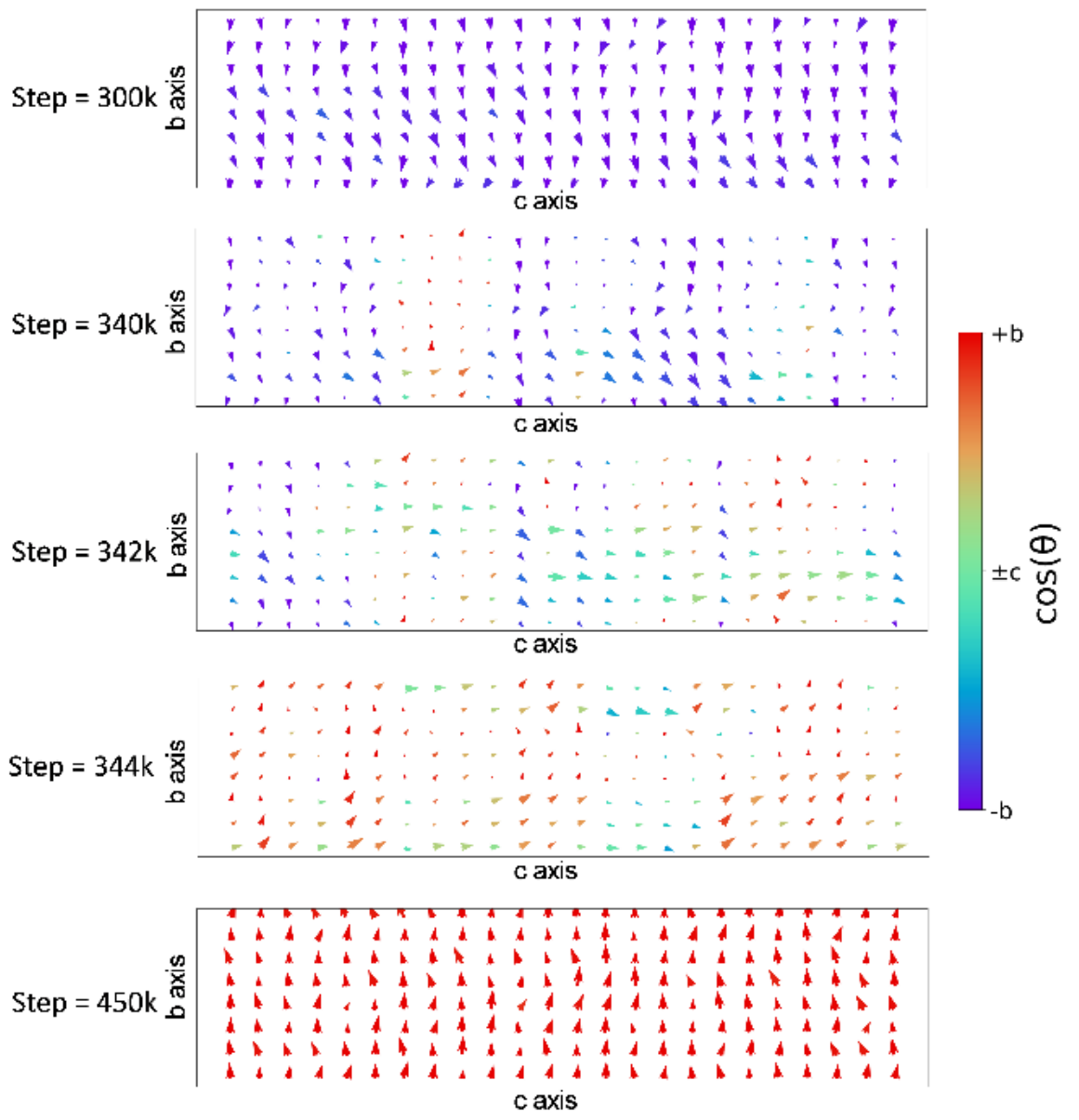

**Figure 7** Evolution of Ti displacement vectors in the 8 × 8 × 24 supercell at selected MD steps under a triangular electric field with a sweep rate of 1 kV/(cm*ps) applied along the b axis at 250 K.

Furthermore, when the electric-field sweep rate is increased to 2 kV/(cm*ps) at the same temperature, DW-mediated switching is recovered, as shown in Figure 8. This behavior can be attributed to the enhanced nucleation and growth rate under a rapidly increasing electric field, which allows the transient domain nuclei to develop into stable DWs before being annihilated by thermal fluctuations. These results demonstrate that DW-mediated switching is determined not only by polarization configurational disorder but also by the kinetic competition between DW nucleation/growth and thermally activated annihilation.

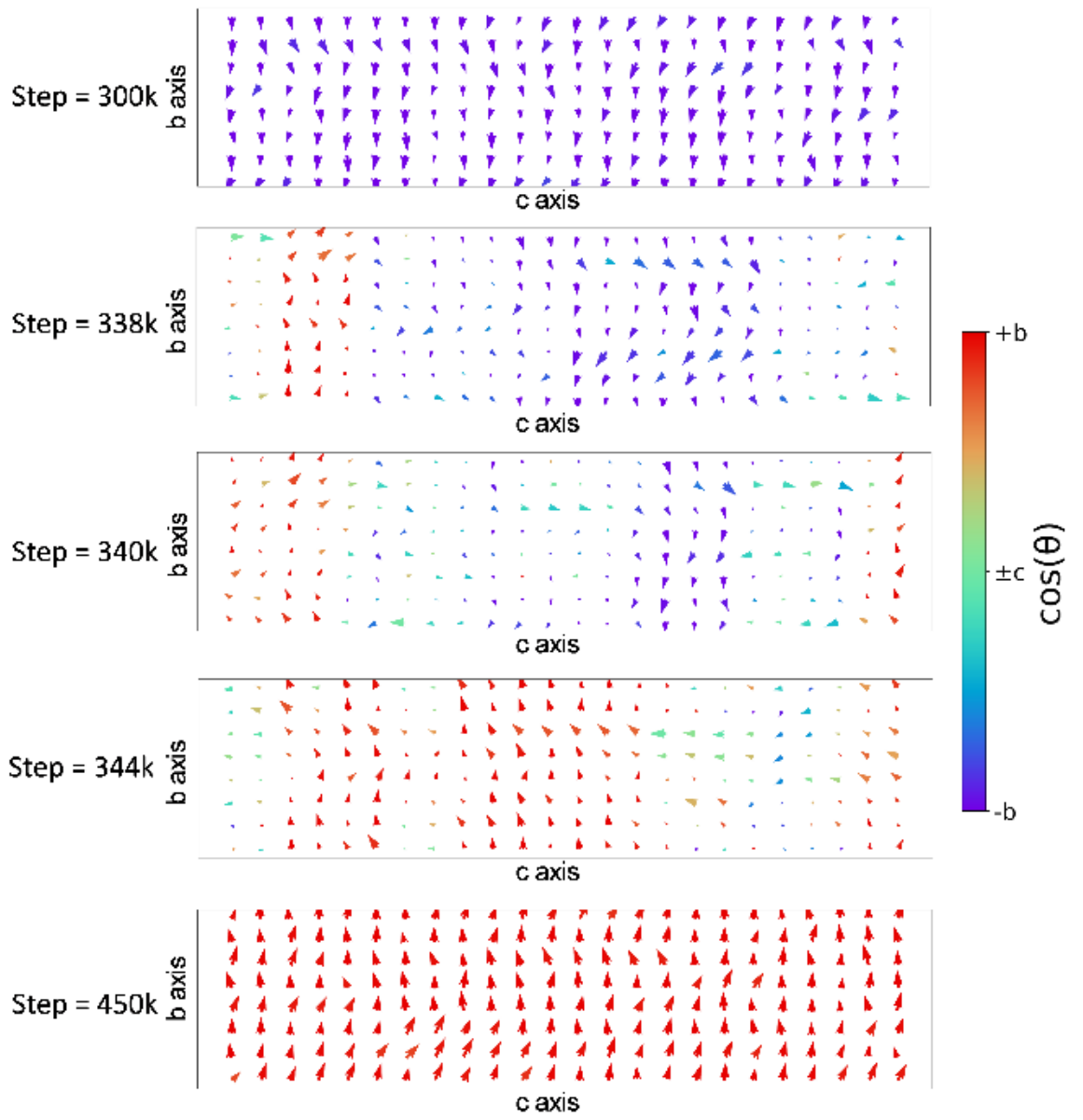

**Figure 8** Evolution of Ti displacement vectors in the 8 × 8 × 24 supercell at selected MD steps under a triangular electric field with a sweep rate of 2 kV/(cm*ps) applied along the b axis at 250 K.

Taken together, these results indicate that neither a large system size nor a high Shannon entropy alone is sufficient to trigger DW-mediated switching. While enhanced polarization configurational disorder facilitates the formation of transient DW nuclei, the operative switching pathway is ultimately determined by the kinetic competition between DW nucleation/growth and thermally activated annihilation. Consequently, DW-mediated switching occurs only when the domain nuclei can develop into stable DWs before being annihilated by thermal fluctuations.

### C. Electric field orientation-dependent P-E hysteresis behavior

To investigate the anisotropic influence of electric-field orientation on polarization switching, we analyzed the P–E hysteresis behavior of the 8×8×32 supercell under a c-axis electric field and compared it with the b-axis electric-field case and the smaller 8×8×8 supercell as shown in Figure 9a. As shown in Figure 9b, the averaged coercive field of the 8×8×32 supercell under the c-axis electric field from three independent simulations is 113.7 ± 7.5 kV/cm, which is substantially larger than that obtained under the b-axis electric field (81.8 ± 3.2 kV/cm). This increase suggests that polarization reversal under the c-axis electric field is associated with a larger effective switching barrier and is no longer dominated by homogeneous switching. Furthermore, the noticeable difference between the b-axis and c-axis electric-field cases indicates that the polarization-switching behavior in the elongated 8×8×32 supercell is strongly anisotropic and depends sensitively on the electric-field orientation relative to the supercell geometry.

*Contact author: poyen@iis.u-tokyo.ac.jp
†Contact author: teru@iis.u-tokyo.ac.jp

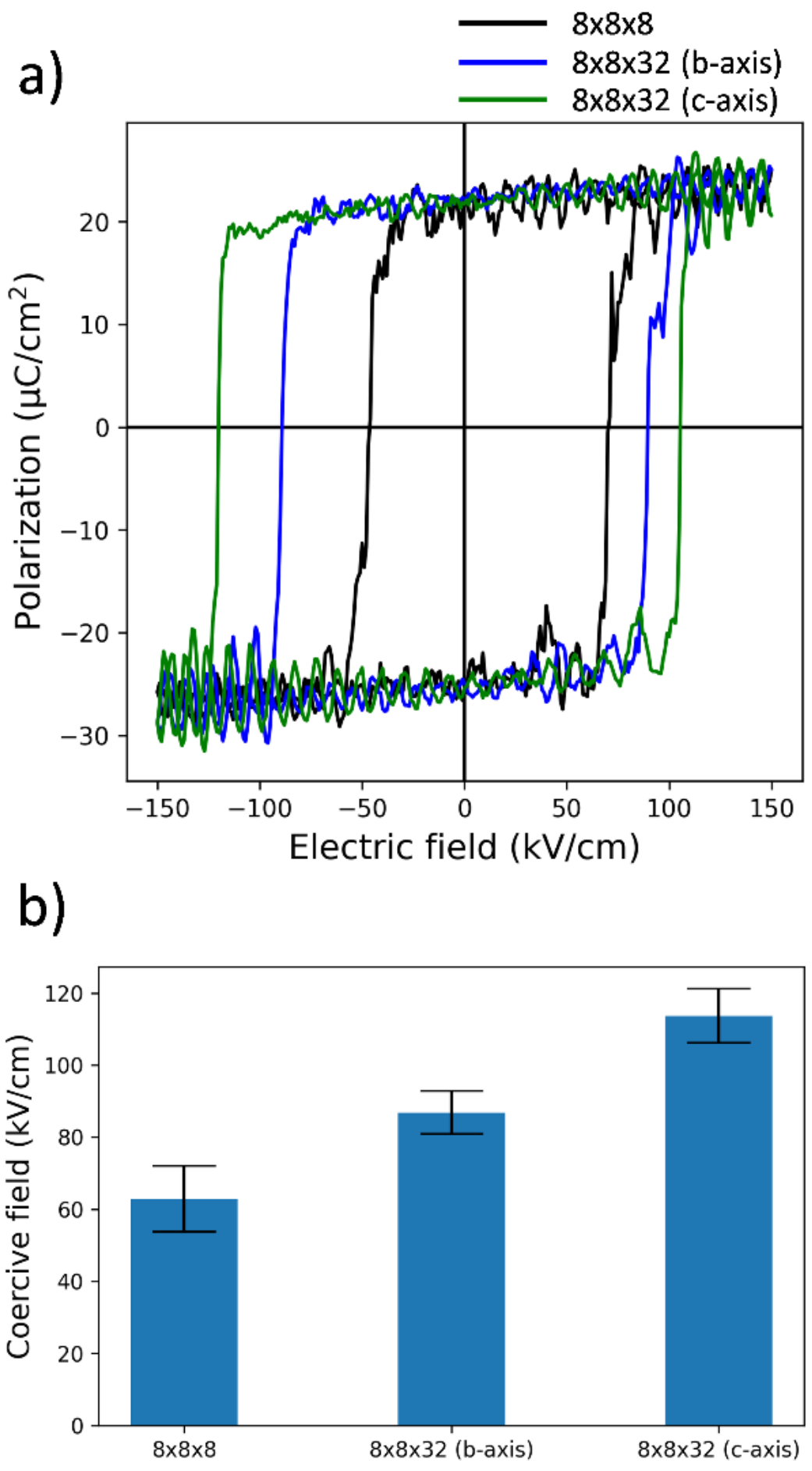

**Figure 9** Electric field orientation-dependent a) hysteresis loops and b) coercive fields of the 8×8×8 and 8×8×32 tetragonal $BaTiO_3$ supercell at 250 K. The black, blue, and green lines represent the hysteresis loop of the 8×8×8 supercell, the 8×8×32 supercell with the b-axis electric field, and the 8×8×32 supercell with the c-axis electric field.

To compare the differences in switching behavior between the b-axis and c-axis electric fields, we further analyzed the evolution of local Ti displacements projected onto the bc plane during polarization reversal in the 8×8×32 supercell under the c-axis electric field, as shown in Figure 10. Initially, without an external electric field (step 0), the local Ti displacements are predominantly aligned along the +c direction, accompanied by slight fluctuations toward the b direction. The polarization reversal process under the c-axis electric field is characterized by enhanced lateral polarization fluctuations and transient formation of intermediate DW-like configurations near the coercive field (step 118k). As the opposite electric field approaches the coercive field, fluctuations along the b axis become significantly enhanced, accompanied by the emergence of local polarization regions oriented close to the ⟨011⟩ and ⟨$01\bar{1}$⟩ directions. These intermediate polarization features resemble transient 90° DW-like structures and indicate the development of local polarization inhomogeneity during switching. Near the coercive field, the local polarizations collectively rotate toward the ±b directions together with the formation of a clear 180° DW (step 119k), followed by rapid reorientation toward the −c direction after switching (step 120k). These observations demonstrate that polarization reversal under the c-axis electric field proceeds through a DW-mediated switching pathway rather than homogeneous collective switching.

Notably, although both the b-axis and c-axis electric fields exhibit 180° DW formation during switching, their microscopic switching behaviors are substantially different. Under the c-axis electric field, the polarization rotation occurs more collectively near the coercive field, whereas under the b-axis electric field, the domain growth process is gradual. We suggest that this difference may originate from finite-size effects associated with the simulation cell geometry. Because the a and b directions contain only eight unit cells, the periodic boundary conditions impose stronger lateral constraints on the formation and propagation of DWs along these directions. As a result, the system tends to preserve the c-axis polarization character while forming intermediate 90° DW-like configurations during switching. In contrast, the elongated c-axis direction in the 8×8×32 supercell allows greater spatial polarization variation and facilitates DW-mediated switching. However, because the polarization rotation under the c-axis electric field occurs more collectively rather than through gradual domain growth, the switching process may require overcoming a larger effective energy barrier. This behavior is consistent with the larger coercive field observed under the c-axis electric field compared with the b-axis electric field.

*Contact author: poyen@iis.u-tokyo.ac.jp
†Contact author: teru@iis.u-tokyo.ac.jp

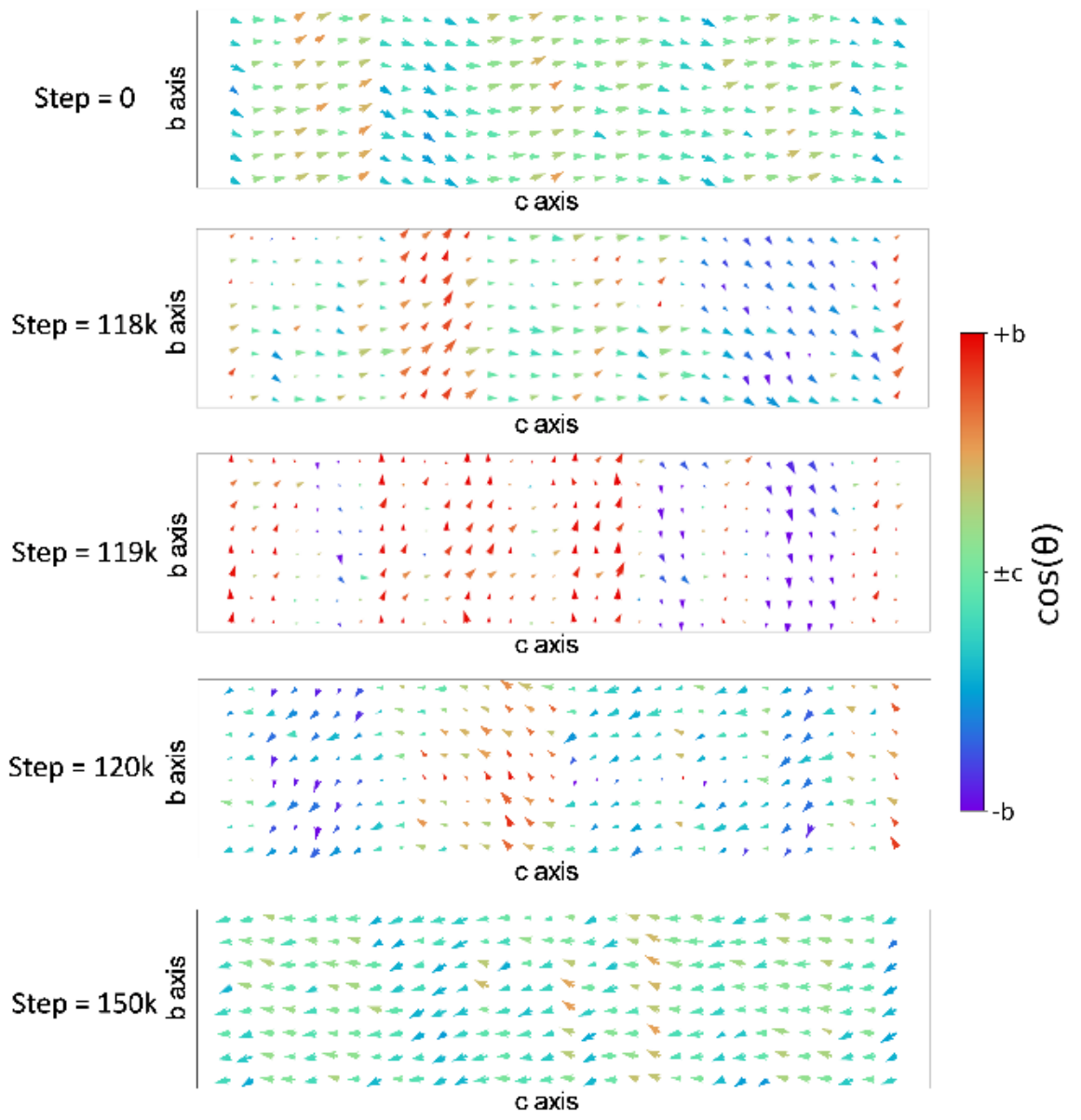

**Figure 10** Evolution of Ti displacement vectors in the 8×8×32 supercell at selected MD steps under a triangular electric field (maximum field: 150 kV/cm) along the c-axis at 250 K. The arrows show the Ti-displacements on each Ti atom.

### D. Stress-dependent P-E hysteresis loop

To further investigate the electromechanical effects in large supercells, we applied uniaxial compressive stress to the 8×8×32 supercell and analyzed its influence on the polarization-switching behavior under electric fields applied along the b and c axes. We first considered the case in which both the uniaxial compressive stress and the triangular electric field were applied along the b axis. Based on our previous studies[11], the critical compressive stress required to induce polarization rotation from the c-axis to the a/b-axis is approximately 120 MPa. To investigate the influence of uniaxial compressive stress below and above this critical value, stresses of 80 and 160 MPa were employed in this study. The resulting P–E hysteresis loops are presented in Figure 11. As shown in the figure, double hysteresis loops are observed under uniaxial compressive stresses of 80 and 160 MPa, consistent with our previous stress-related investigations[11]. As discussed in our earlier study, compressive stress parallel to the switching direction energetically favors polarization states perpendicular to the electric field, thereby stabilizing intermediate polarization configurations during the switching process and producing characteristic double hysteresis behavior.

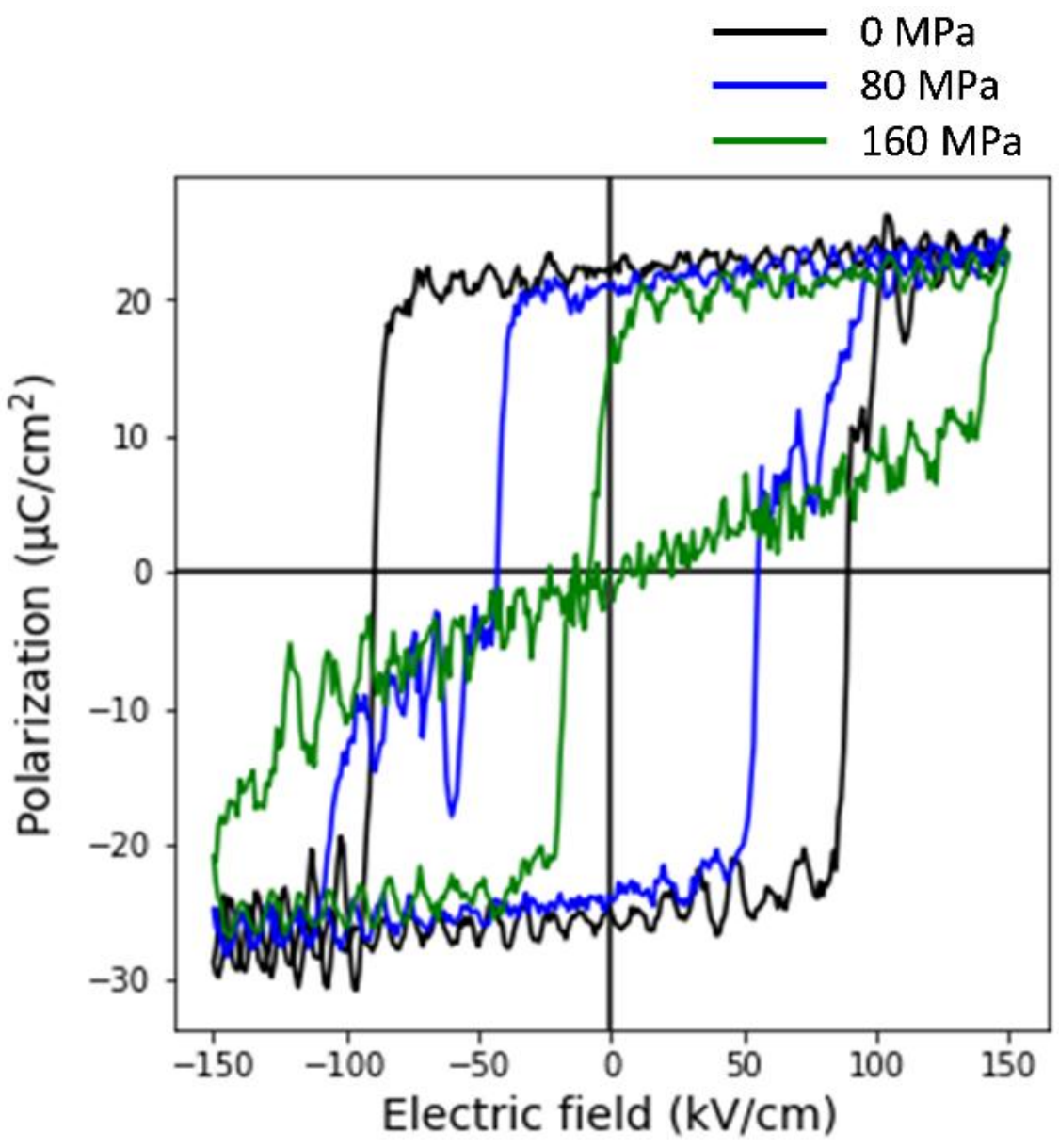

**Figure 11** The P–E hysteresis loops under the *b*-axis electric field for the 8×8×32 tetragonal $BaTiO_3$ supercell at 250 K under different *b*-axis compressive stresses. The black, blue, and green curves correspond to compressive stresses of 0, 80, and 160 MPa, respectively.

Compared with the smaller supercell case examined in our earlier study[11], the 8×8×32 supercell exhibits a noticeably larger coercive field during the first polarization-switching process, suggesting the presence of a larger effective switching barrier associated with the large-supercell switching mechanism. Furthermore, unlike the smaller-supercell case, where a paraelectric-like P–E response emerged under high compressive stress, the large supercell still maintains a double hysteresis loop even at 160 MPa. This difference may originate from the larger electric-field amplitude employed in the present study, which enables polarization reversal to occur despite the mechanically stabilized intermediate polarization states. In addition, the electric-field region corresponding to the intermediate polarization state becomes noticeably wider under 160 MPa than under 80 MPa, indicating that stronger compressive stress further stabilizes the metastable polarization configuration and enlarges the metastable switching region. This behavior suggests that the competition between electric-field-driven polarization reversal and stress-induced stabilization of perpendicular polarization states becomes increasingly pronounced under stronger compressive stress. As a result, the system remains trapped within intermediate polarization configurations over a broader electric-field range before complete polarization reversal occurs.

*Contact author: poyen@iis.u-tokyo.ac.jp
†Contact author: teru@iis.u-tokyo.ac.jp

To further clarify the orientation dependence of this electromechanical coupling, we next considered the case in which the compressive stress was applied perpendicular to the polarization-switching direction. In this configuration, uniaxial compressive stress was applied along the c axis, whereas the triangular electric field was applied along the b axis in the 8×8×32 supercell. In contrast to the parallel-field case, the hysteresis loops remain nearly unchanged even under compressive stress up to 160 MPa as shown in Figure 12. Neither double hysteresis behavior nor a significant change in coercive field is observed, indicating that compressive stress applied perpendicular to the switching direction has only a limited influence on the polarization-switching dynamics. These results demonstrate that the influence of uniaxial compressive stress on polarization switching strongly depends on its relative orientation with respect to the electric-field direction. When the compressive stress is applied parallel to the switching direction, intermediate polarization states become energetically stabilized, substantially modifying the switching pathway and hysteresis behavior. In contrast, compressive stress applied perpendicular to the switching direction couples only weakly to the polarization reversal process and therefore produces minimal changes in the ferroelectric response.

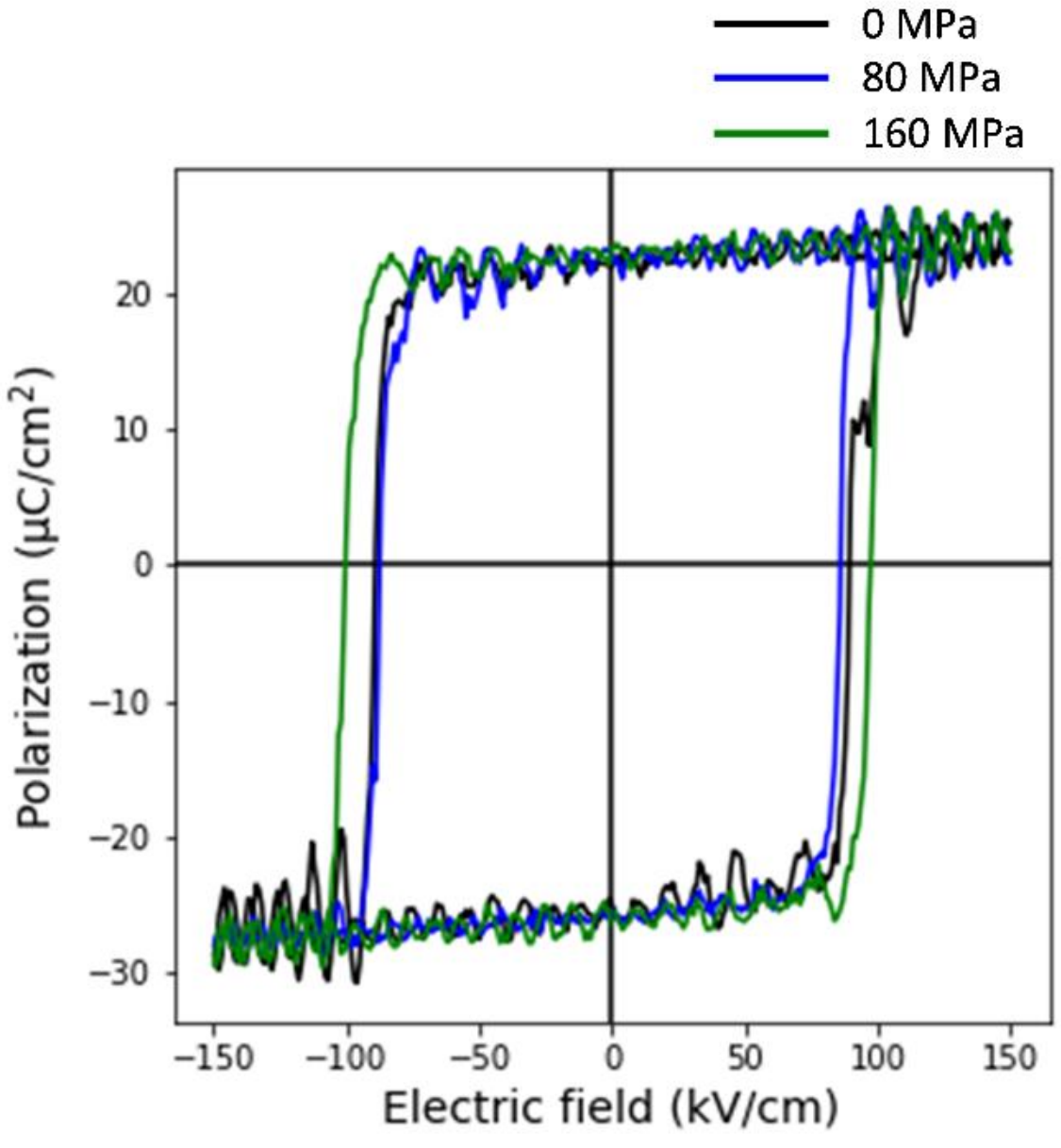

**Figure 12** The P–E hysteresis loops under the *b*-axis electric field for the 8×8×32 tetragonal $BaTiO_3$ supercell at 250 K under different *c*-axis compressive stresses. The black, blue, and green curves correspond to compressive stresses of 0, 80, and 160 MPa, respectively.

Based on the above observations, the influence of uniaxial compressive stress on the hysteresis behavior is strongly dependent on its orientation relative to the electric-field direction. In the case where both the uniaxial compressive stress and the triangular electric field are applied along the c axis of the 8×8×32 supercell, the emergence of double hysteresis behavior becomes significantly suppressed compared with the b-axis electric-field case, as shown in Figure 13. This behavior suggests that increasing the supercell dimension along the stress and electric-field direction weakens the effective constraints imposed by periodic boundary conditions, thereby enhancing local polarization fluctuations and configurational freedom within the system. As a result, the stress-induced stabilization of polarization states perpendicular to the electric-field direction becomes partially relaxed, making the formation of intermediate polarization states more difficult.

Consistent with this interpretation, the system retains a single hysteresis loop under compressive stresses of 80 and 120 MPa, despite a substantial reduction in the coercive field. Double hysteresis behavior emerges only when the compressive stress reaches 160 MPa, whereas smaller supercells and the b-axis electric-field case exhibit double hysteresis loops already at 80 MPa. These results indicate that the critical stress required to stabilize intermediate polarization states strongly depends on the accessible length scale along the stress and electric-field direction. Nevertheless, sufficiently large compressive stress can still overcome polarization fluctuations and eventually stabilize polarization states perpendicular to the electric-field direction, leading to the emergence of double hysteresis behavior.

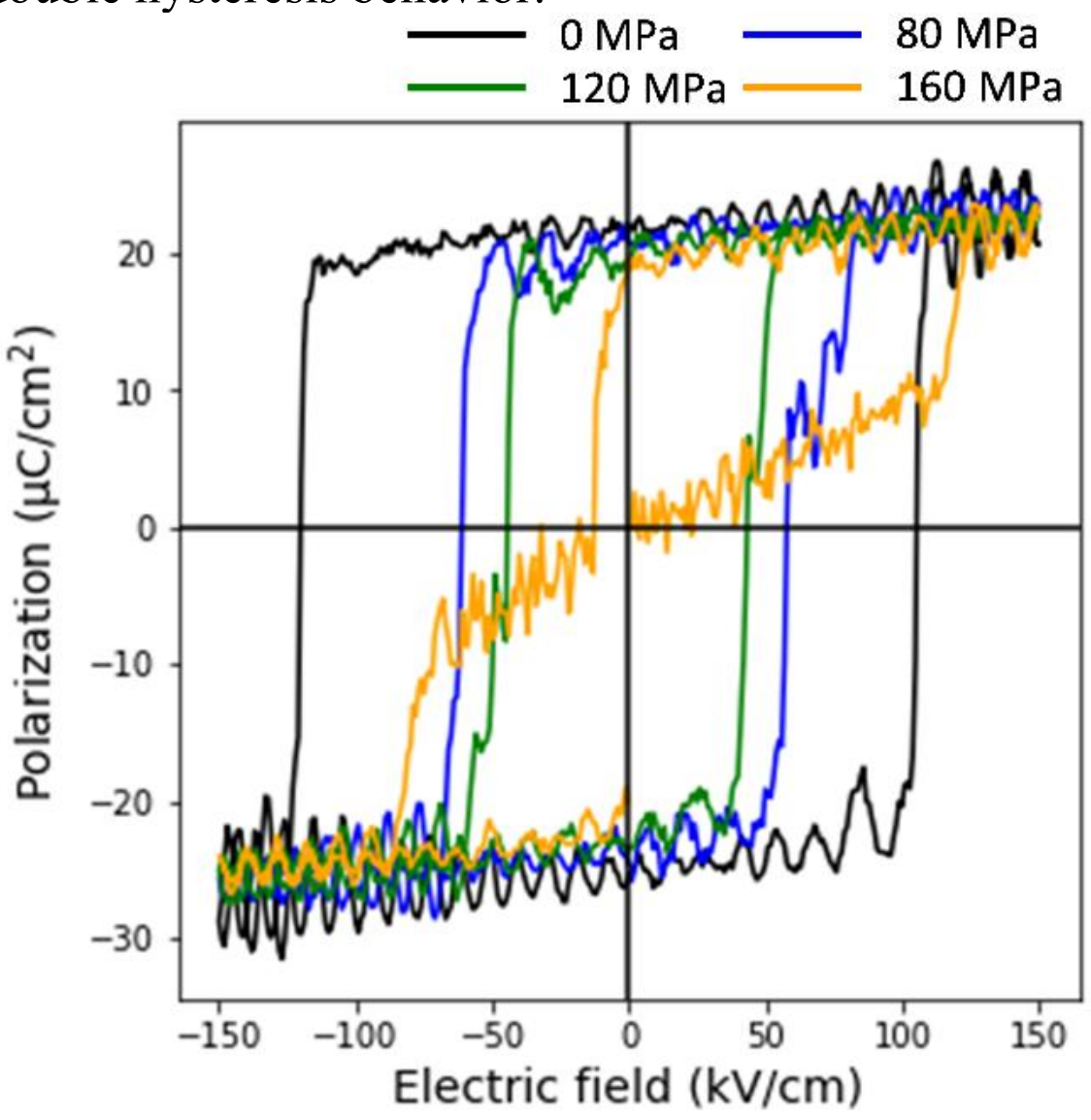

**Figure 13** The P–E hysteresis loops under the *c*-axis electric field for the 8×8×32 tetragonal $BaTiO_3$

*Contact author: poyen@iis.u-tokyo.ac.jp
†Contact author: teru@iis.u-tokyo.ac.jp

supercell at 250 K under different *c*-axis compressive stresses. The black, blue, green, and orange curves correspond to compressive stresses of 0, 80, 120, and 160 MPa, respectively.

Taken together, the results reveal a coherent physical picture of size-dependent polarization switching in $BaTiO_3$. In small supercells, suppressed polarization fluctuations inhibit the formation of stable DW nuclei, leading to nearly homogeneous polarization switching with relatively low coercive fields. As the supercell size increases, enhanced polarization fluctuation, quantified by the Shannon entropy, promotes the formation of transient DW nuclei and shifts the switching behavior toward a DW-mediated mechanism accompanied by significantly higher coercive fields. However, the additional temperature-dependent simulations demonstrate that solely high Shannon entropy might not be sufficient to process DW-mediated switching. Instead, the kinetic competition between DW nucleation/growth and the thermal DW annihilation also plays an important role in determining the polarization switching behavior.

This size-dependent transition is also reflected in the electromechanical response. The critical stress required to stabilize intermediate polarization states and induce double hysteresis behavior increases with supercell size, while the switching pathway becomes increasingly sensitive to the relative orientation between the applied electric field and mechanical stress. These results demonstrate that system size is not merely a computational parameter but an intrinsic physical variable that controls the probability of stable DW formation through its influence on polarization fluctuations and switching kinetics. It should be noted, however, that these conclusions are drawn within the nanosecond timescale accessible to molecular dynamics simulations. At experimentally relevant timescales[17], which are many orders of magnitude longer, additional thermally activated switching pathways and kinetic processes may become important. Therefore, while the present simulations provide atomistic insight into the intrinsic switching mechanisms at short timescales, caution should be exercised when directly extrapolating these results to long-timescale experimental switching dynamics.

## IV. CONCLUSIONS

In conclusion, this study demonstrates that polarization switching in tetragonal $BaTiO_3$ occurs through two distinct physical regimes—nearly homogeneous switching and DW-mediated switching—and that system size plays a central role in determining the accessible switching pathway. Using machine-learning-potential molecular dynamics based on the MACEField model, we show that increasing the supercell size enhances polarization configurational disorder, quantified by the Shannon entropy, thereby facilitating the formation of transient 180° DW nuclei and promoting the emergence of DW-mediated switching. However, our additional analyses indicate that neither a large system size nor a high Shannon entropy alone is sufficient to determine the switching pathway. Instead, the switching behavior is also influenced by the kinetic competition between DW nucleation/growth and annihilation. Consequently, the coercive field increases by more than 50% as the switching mechanism evolves from homogeneous polarization reversal to a DW-mediated process.

Furthermore, the switching pathway and electromechanical response are strongly influenced by both the supercell geometry and the relative orientation of the applied electric field and mechanical stress, with the critical stress required for double-hysteresis behavior increasing significantly with system size. These results suggest that finite-size effects and switching kinetics together govern the stability of DW nuclei and the resulting polarization-switching pathway. Overall, these findings demonstrate that finite-size effects are not merely numerical artifacts but physically meaningful factors governing DW dynamics and polarization-switching behavior in ferroelectric $BaTiO_3$.

From a practical perspective, our results highlight that atomistic simulations of ferroelectric switching should carefully consider system size and the associated switching kinetics to correctly capture the operative switching mechanism. Future work extending this framework to experimentally relevant timescales, defect-containing systems, and other ferroelectric materials will further clarify the generality of the size- and kinetic-dependent switching mechanism identified in this study.

## ACKNOWLEDGMENTS

This study was supported by the Ministry of Education, Culture, Sports, Science and Technology (MEXT) (Nos. 26K01205, 26K22608, 25K24645), and the New Energy and Industrial Technology Development Organization (NEDO). PYC would acknowledge the support of JST SPRING (Grant Number JPMJSP2108).

## AUTHOR CONTRIBUTIONS

Po-Yen Chen conceptualized the work, conducted the investigation and data analysis, developed the methodology and software, visualized the results, and wrote the original draft of the manuscript. Teruyasu Mizoguchi supervised the project, acquired funding, and reviewed and edited the manuscript.

*Contact author: poyen@iis.u-tokyo.ac.jp
†Contact author: teru@iis.u-tokyo.ac.jp

## DATA AVAILABILITY

All models utilized in this work, including the MACE and MACEField models, are publicly available in our GitHub repository, Finetuned-MACE-model (https://github.com/nmdl-mizo/Finetuned-MACE-model.git).

## APPENDIX A: MaceField Model performance

The MACEfield model[27] employed in this study was developed based on our previously finetuned MACE model[26] and further trained using a BEC dataset[11] containing 2,038 solid-state $BaTiO_3$ structures to enable BEC prediction. The energy and atomic-force prediction accuracies of the MACEfield model are compared with those of the original MACE model in Table 1. As shown in the table, the prediction accuracies of the MACE and MACEfield models remain highly similar, with only minor improvements observed for the MACEfield model due to the additional finetuning using the solid-state dataset. These results indicate that the introduction of the BEC training does not significantly affect the energy and force prediction performance of the original MACE model. Therefore, to maintain consistency with our previous studies, the original finetuned MACE model was retained as the MLP in the present MLP-MD framework, while the MACEfield model was used specifically for BEC prediction.

Table 1 The energy and force prediction ability of the MACE model and the MACEField model.

| MAE | MACE | MACEField |
|---|---|---|
| Energy (meV/atom) | 0.85 | 0.43 |
| Force (meV/Å) | 17.5 | 14.9 |

In addition, we examined the phonon dispersions of the 9×9×9 $BaTiO_3$ supercell to evaluate whether the MACEField model can reproduce long-range interaction effects, as shown in Figure 14. Four models were considered in this comparison: the original MACE model, the MACEField model finetuned from the MACE model, the MACELES model[43-47], and the MACEField model finetuned from the MACELES model. As shown in the figure, the MACEField models do not reproduce the characteristic phonon behavior near the Γ point associated with long-range electrostatic interactions, even when finetuned based on the MACELES model. These results indicate that the present MACEField framework does not explicitly capture long-range interaction effects. Nevertheless, as discussed in our previous study, long-range interactions mainly influence the quantitative properties of bulk $BaTiO_3$, whereas the qualitative polarization-switching behavior remains largely unchanged. Therefore, the main discussions and conclusions regarding polarization-switching mechanisms in the present study are not expected to be significantly affected by the absence of explicit long-range interactions.

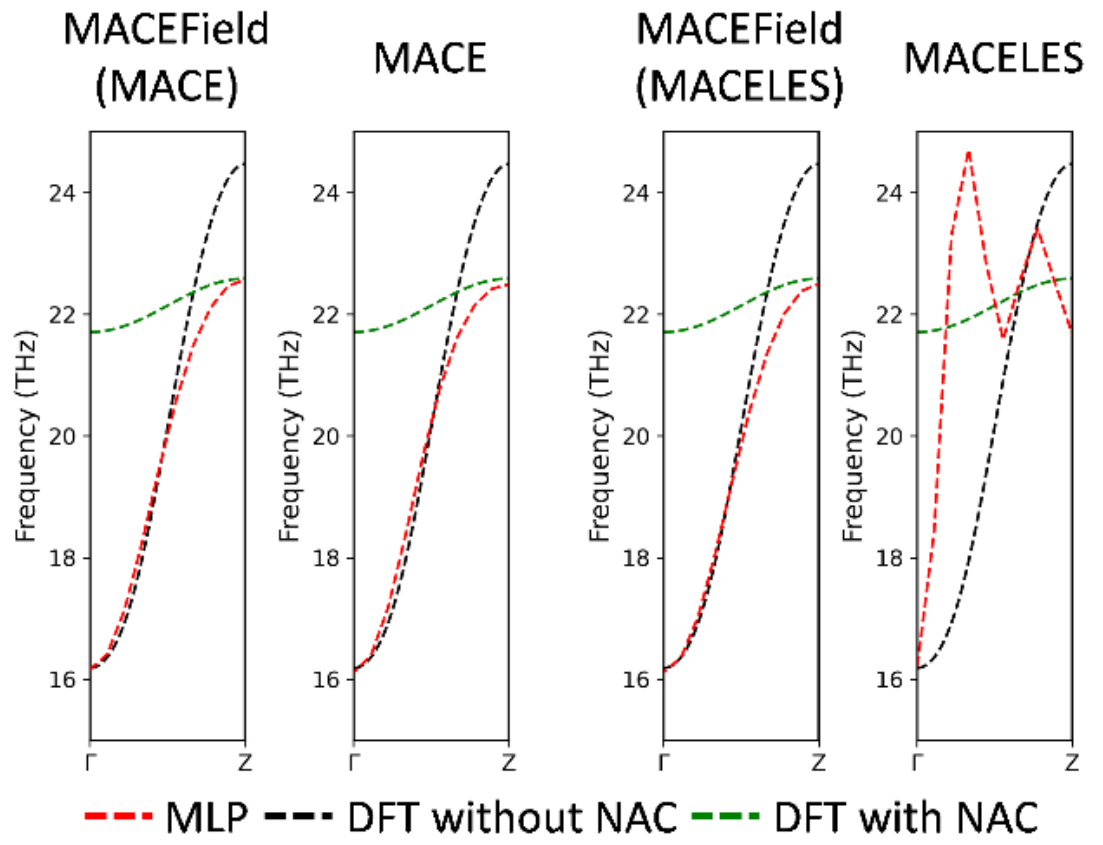

**Figure 14** The phonon dispersion on the Γ-Z path calculated by the MACEField finetuned by MACE, the MACE, the MACEField finetuned by MACELES, and the MACELES model. The red, black, and green dashed lines show the results obtained by MLP, DFT without non-analytic correction (NAC), and DFT with NAC, respectively.

Next, to evaluate the performance of the MACEField model in the present workflow, we investigated its BEC prediction accuracy and compared it with the previously developed Equivar model[28, 29]. The comparison results are summarized in Table 2. As shown in the table, the MACEField model exhibits improved prediction accuracy for solid-state $BaTiO_3$ structures compared with the Equivar model. This improvement is mainly attributed to the use of a solid-state-focused training dataset in the present MACEField model, whereas the Equivar model was additionally trained using liquid structures. In contrast, for liquid structures, the Equivar model shows better prediction performance than the MACEField model.

Table 2 The prediction ability of the Equivar model and the MACEField model for solid and liquid structures.

| MAE (e) | Equivar model | MACEField |
|---|---|---|
| Solid | 0.03 | 0.01 |
| Liquid | 0.07 | 0.26 |

To further validate the applicability of the MACEField model in electric-field MD simulations,

*Contact author: poyen@iis.u-tokyo.ac.jp
†Contact author: teru@iis.u-tokyo.ac.jp

we compared the P–E hysteresis loops obtained using the previous framework (MACE + Equivar)[11, 25] and the present framework (MACE + MACEField) for the 8×8×8 $BaTiO_3$ supercell. As shown in Figure 15, the hysteresis loops obtained using the two methods remain similar, indicating that the replacement of the Equivar model by the MACEField model does not significantly affect the polarization-switching behavior observed in our previous studies.

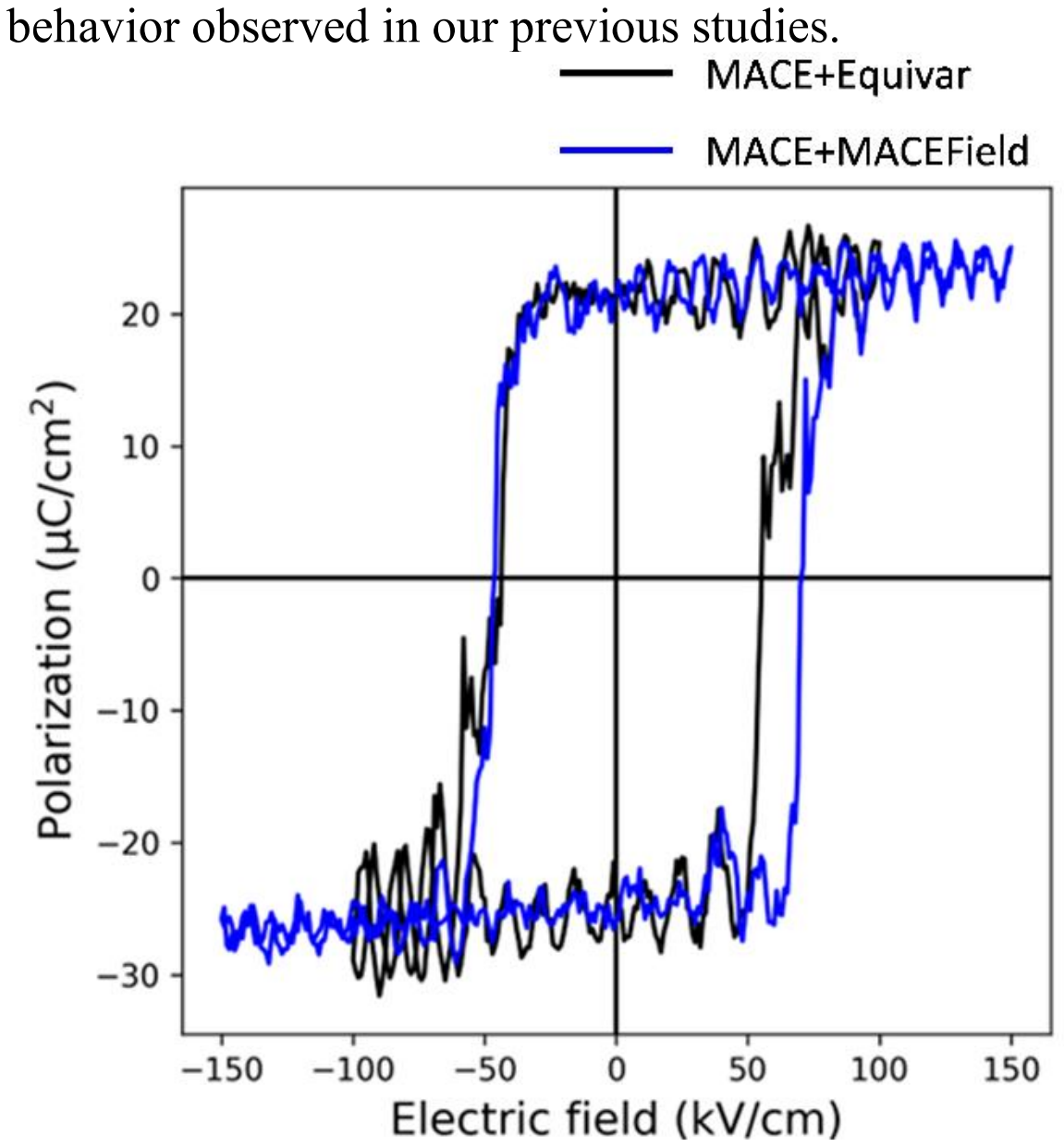

**Figure 15** The hysteresis loop obtained by the MACE + Equivar model (black) and the MACE + MACEField model (blue). The data of MACE +Equivar model was previously reported from Ref. [43].

The computational performance of the Equivar model and MACEField was further compared using 20k-step MD simulations for $BaTiO_3$ supercells with different sizes, as summarized in Figure 16. For the previous framework, the calculation time per atom per MD step gradually increases with increasing system size. In contrast, the MACEField framework maintains nearly constant computational cost as the supercell size increases, demonstrating significantly improved computational scalability for large systems. Furthermore, the application of cuEquivariance acceleration[38] provides additional improvement in computational efficiency.

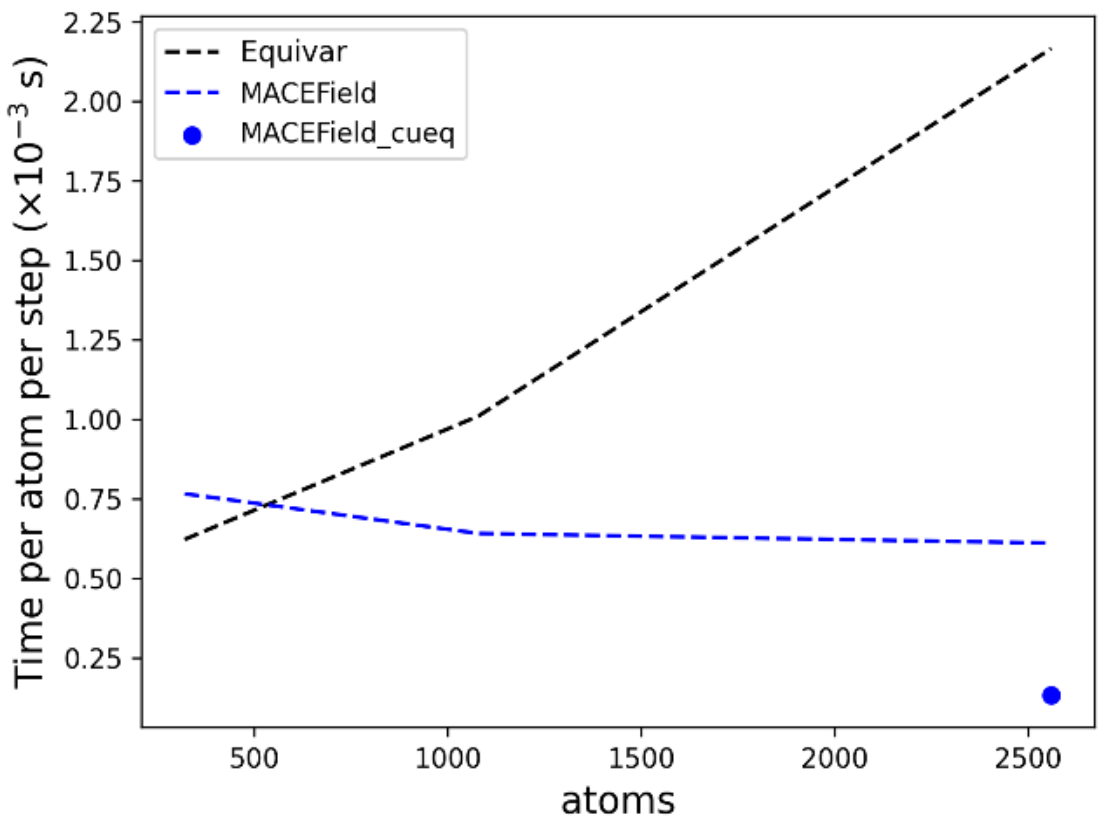

**Figure 16** Number of atom-dependent calculation time. The black dashed, blue dashed, and blue point represent the Equivar model, the MACEField model, and the MACEField model with cuEquivariance acceleration.

The GPU memory usage during MD simulations is compared in Figure 17. Without acceleration, the MACEField framework requires approximately five times larger GPU memory usage than the original MACE model, reflecting the increased computational cost associated with the BEC prediction. However, after applying cuEquivariance acceleration, the GPU memory usage of the MACEField framework is significantly reduced and becomes even lower than that of the original MACE model without acceleration. These results demonstrate that the MACEField framework can provide both high prediction accuracy and high computational efficiency for electric-field MD simulations of large-scale ferroelectric systems. Therefore, the MACEField model was adopted in the present study.

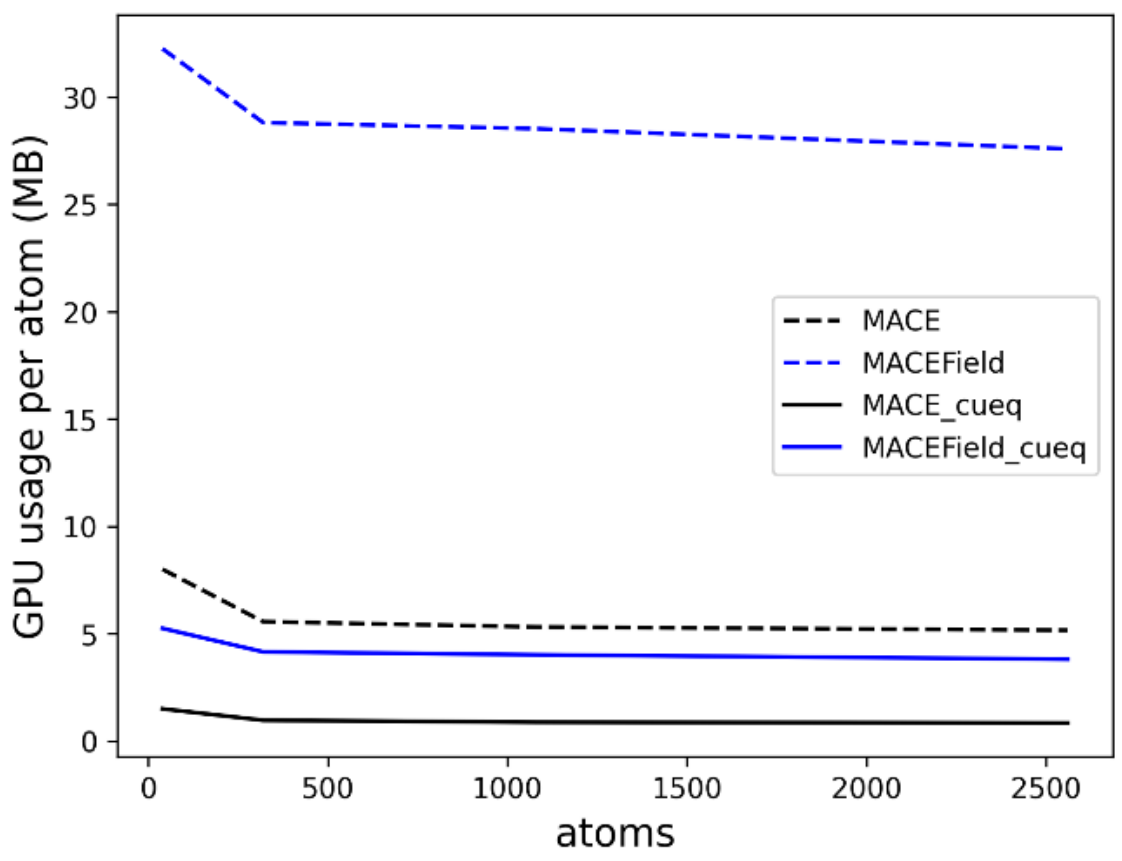

**Figure 17** Number of atom-dependent GPU usage by MACE (black dashed), MACEField (blue dashed), MACE with cuEquivariance acceleration, and MACEField with cuEquivariance acceleration.

*Contact author: poyen@iis.u-tokyo.ac.jp
†Contact author: teru@iis.u-tokyo.ac.jp

**APPENDIX B: Electric field sweep rate-dependent hysteresis behavior**

Since the coercive field is strongly affected by the electric-field sweep rate, we further investigated the sweep rate dependence of the hysteresis loop by varying the electric field sweep rate from 0.5 kV/(cm*ps) to 6 kV/(cm*ps) for the 8×8×24 supercell, corresponding to electric-field sweep rates ranging from 6 kV/(cm*ps) to 0.5 kV/(cm*ps). The results are presented in Figure 18a. As shown in Figure 18a, simulations with larger sweep rate (6 kV/(cm*ps) and 4 kV/(cm*ps)) exhibit pronounced fluctuations in the polarization response. In particular, the coercive field obtained from the 6 kV/(cm*ps) simulation shows large statistical variations, making it unsuitable for identifying the overall trend. Therefore, the analysis of the sweep rate dependence is based on the results from 0.5 kV/(cm*ps) to 4 kV/(cm*ps), as summarized in Figure 18b.

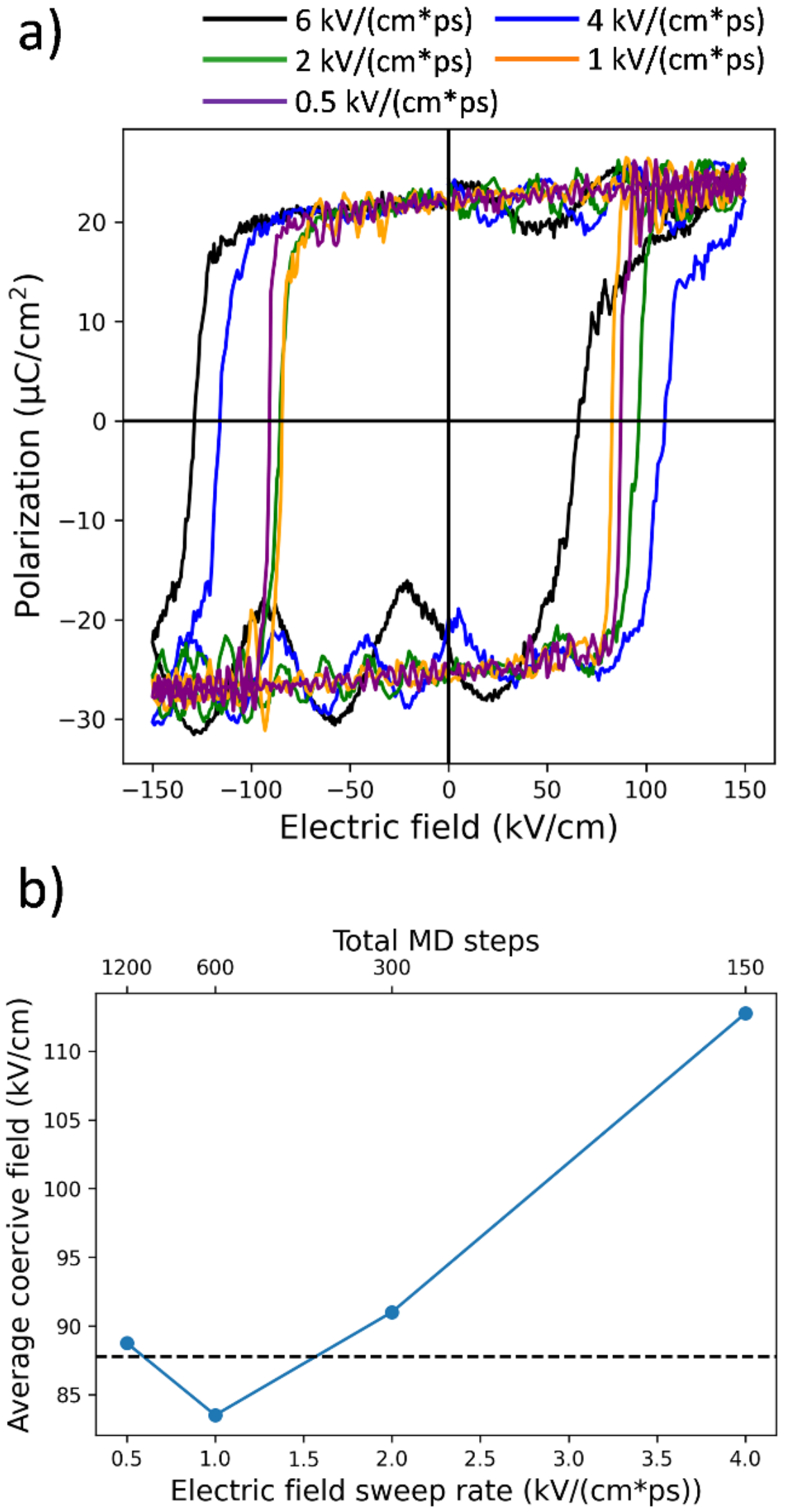

**Figure 18** Rate dependence of (a) hysteresis loops and (b) coercive field for the 8 × 8 × 24 $BaTiO_3$ supercell. The dashed line indicates the average coercive field in the converged sweep rate regime.

As the electric-field sweep rate decreases, the coercive field decreases rapidly and approaches a nearly constant value of approximately 87 kV/cm once the cycle time reaches about 2 kV/(cm*ps). This result indicates that the coercive field is essentially converged with respect to the electric-field sweep rate under the present simulation conditions. Therefore, the 1 kV/(cm*ps) used throughout the main text is sufficiently long to eliminate significant sweep rate effects and provides reliable coercive-field values.

**APPENDIX C: Shannon entropy calculation**

To quantify the degree of polarization fluctuation during the polarization-switching process, the Shannon entropy was calculated from the distribution of local Ti-displacement directions. For each MD step, the Ti-displacement vectors were first normalized into unit vectors to remove the influence of displacement magnitude and retain only the polarization orientation information. The normalized vectors were then transformed into spherical coordinates $(\theta, \phi)$, where $\theta$ represents the polar angle and $\phi$ represents the azimuthal angle. The orientation distribution was discretized using a two-dimensional histogram on the spherical coordinate space with $n_{\mathrm{bins}} \times n_{\mathrm{bins}}$ angular grids. The probability of finding local polarization vectors within each angular region was calculated as

$$p_i = \frac{N_i}{N}$$

where $N_i$ is the number of vectors in the $i$-th angular bin and $N$ is the total number of Ti-displacement vectors. This normalization minimizes the influence of the total number of atoms, enabling direct comparison between different supercell sizes. The Shannon entropy was then evaluated as

$$S = -\sum_i p_i \ln p_i$$

where the summation was performed over all nonzero probability bins. A larger Shannon entropy corresponds to a broader distribution of local polarization orientations, indicating enhanced polarization fluctuation and configurational disorder within the supercell.

To examine the sensitivity of the calculated maximum Shannon entropy during the polarization-switching process to the histogram resolution, convergence tests were performed using the 8×8×32 supercell, as shown in Figure 19. The calculated entropy gradually increases with increasing bin number and becomes essentially converged when the number of bins reaches approximately 150. Therefore, all Shannon entropy values reported in this work were calculated using $n_{\text{bins}}$ = 150.

*Contact author: poyen@iis.u-tokyo.ac.jp
†Contact author: teru@iis.u-tokyo.ac.jp

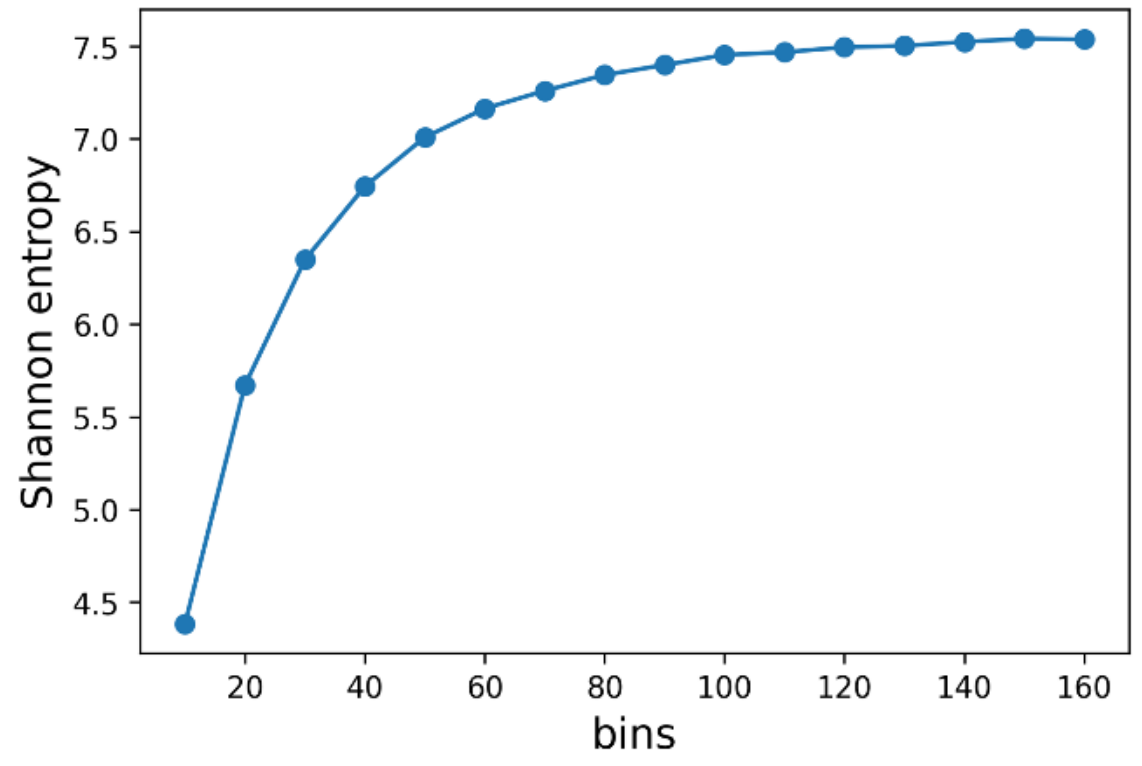

**Figure 19** Dependence of the maximum Shannon entropy on the bin size during the simulation of the 8 × 8 × 32 tetragonal $BaTiO_3$ supercell at 250 K.

The Shannon entropy was evaluated at every MD step throughout the electric-field-driven simulation. Rather than using a time-averaged value, the maximum Shannon entropy during the polarization-switching process was adopted for quantitative comparison among different simulation conditions. Figure 20 compares the time evolution of the Shannon entropy for the 8×8×8 and 8×8×32 supercells under the triangular electric field. Although both systems exhibit a transient increase in Shannon entropy during polarization reversal, the larger supercell reaches a substantially higher maximum entropy, indicating stronger polarization configurational disorder during switching.

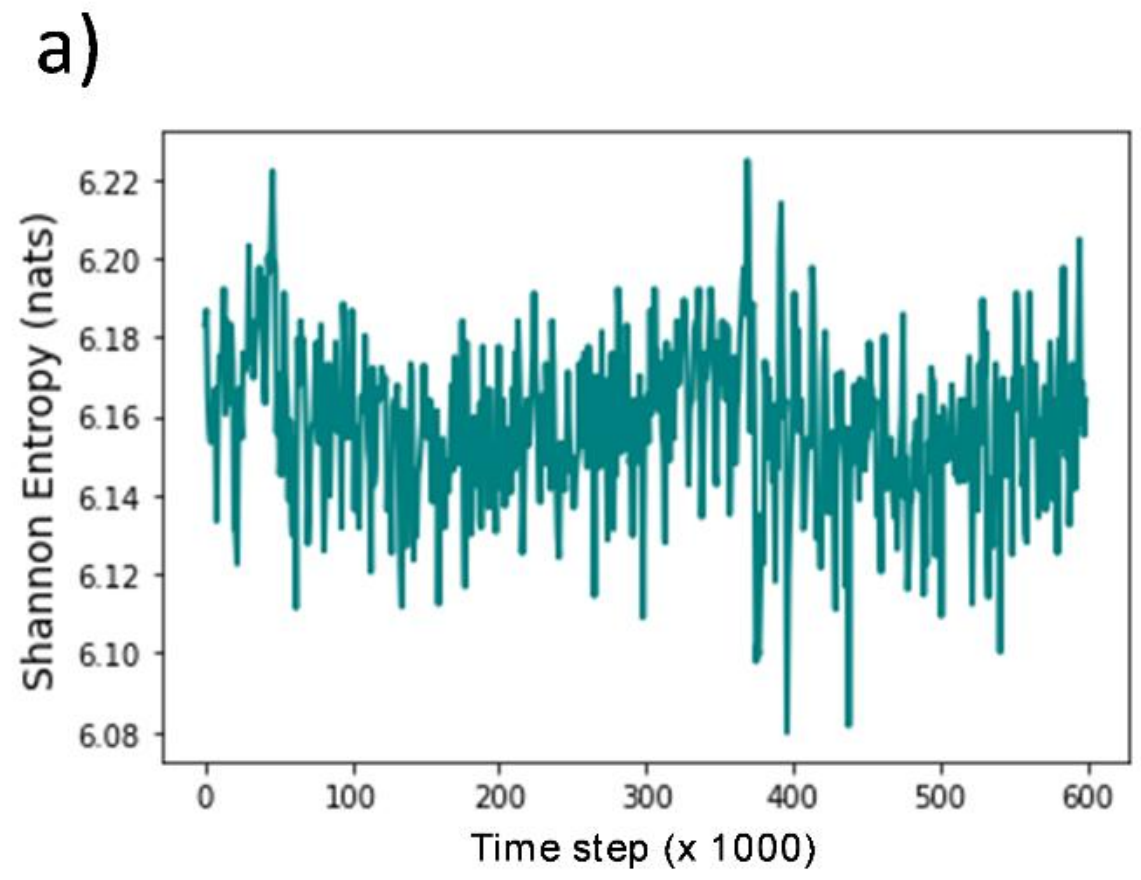

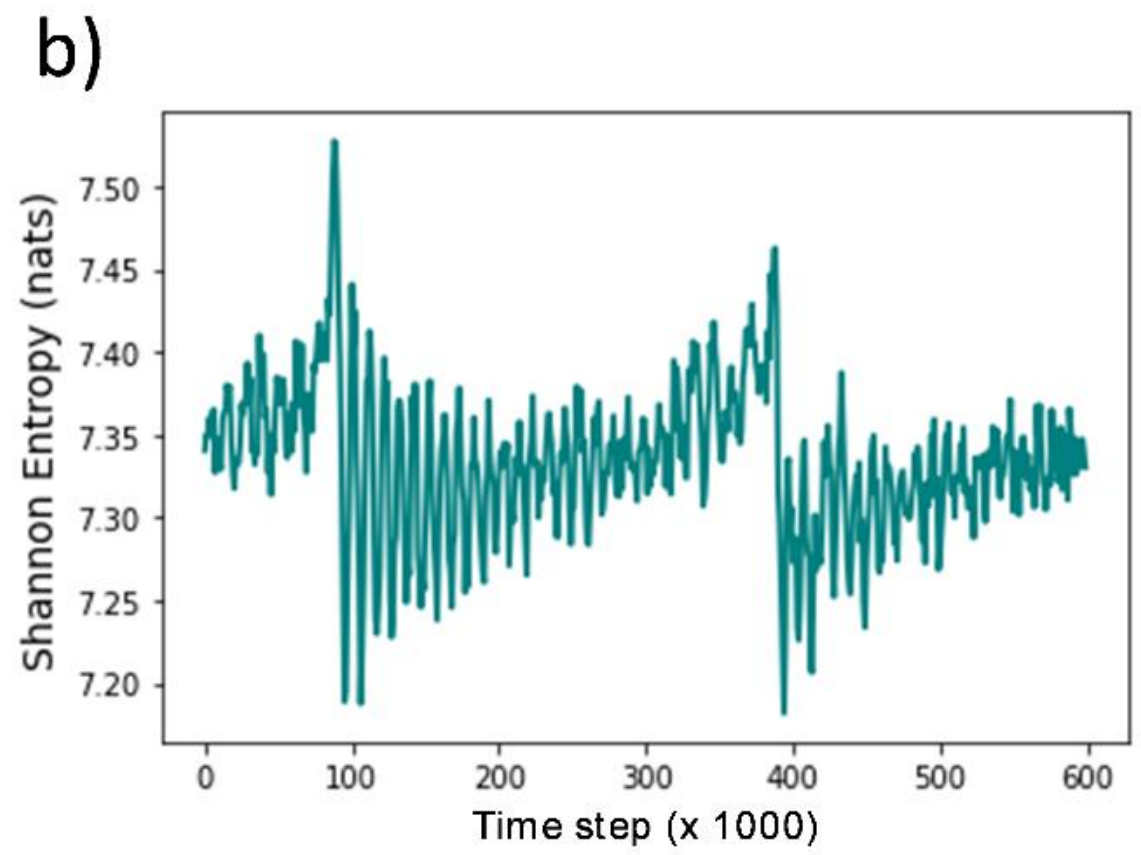

**Figure 20** Step dependence of the Shannon entropy during the MD simulations of the a) 8 × 8 × 8 and b) 8 × 8 × 32 $BaTiO_3$ supercells.

**APPENDIX D: Long range interaction effect on DW-mediated switching**

As discussed in Appendix A, although the MACEField model was developed by fine-tuning the MACELES foundation model, it does not explicitly include long-range electrostatic interactions. To evaluate the influence of long-range electrostatics on DW-mediated polarization switching, additional simulations were performed using the MACELES model[43] as the interatomic potential while retaining the MACEField model for the prediction BECs. The corresponding hysteresis loop is shown in Figure 21, exhibiting qualitatively the same switching behavior as that obtained using the MACEField model, with only a slight difference in the coercive field.

*Contact author: poyen@iis.u-tokyo.ac.jp
†Contact author: teru@iis.u-tokyo.ac.jp

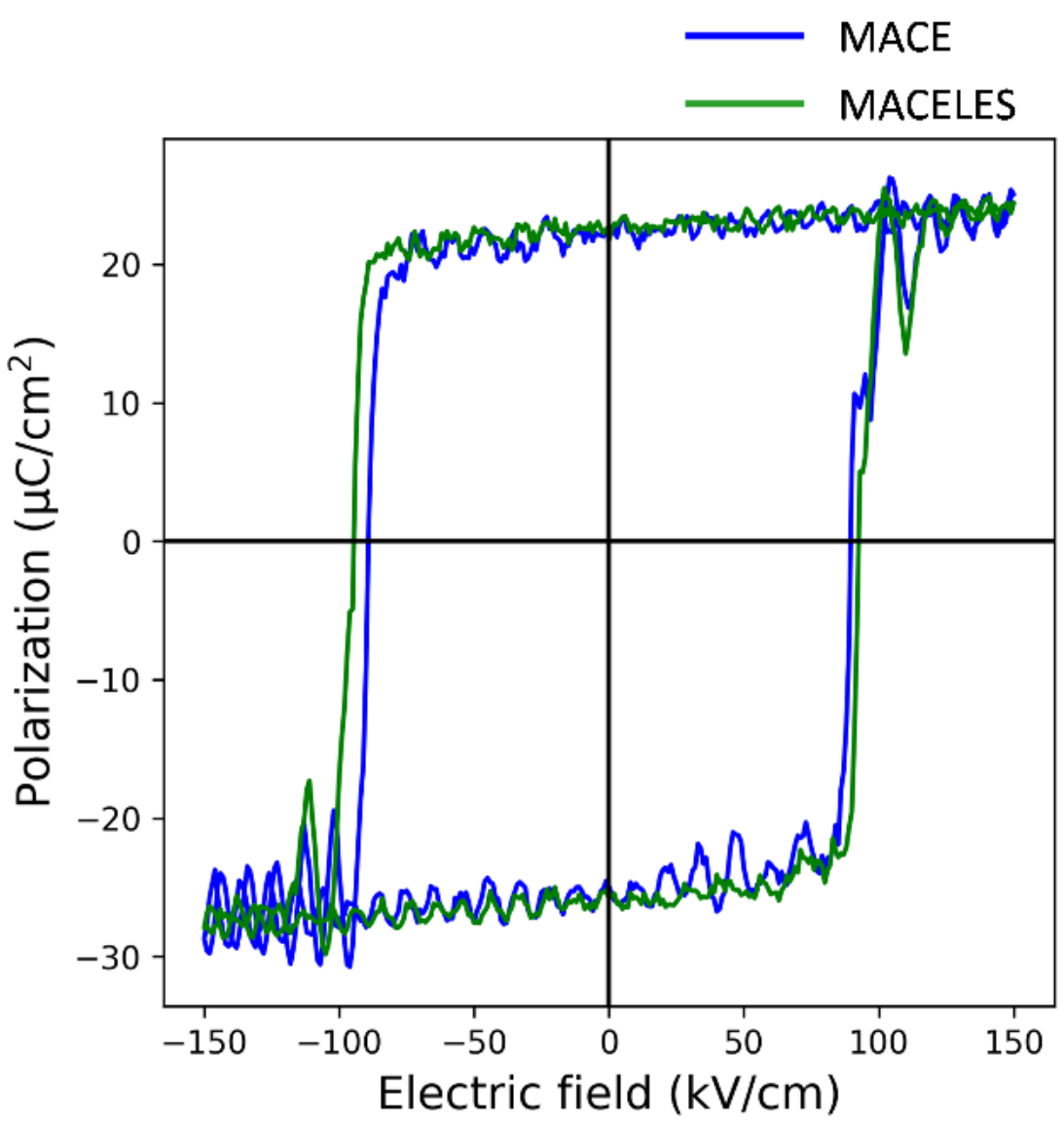

**Figure 21** The hysteresis loop of 8 × 8 × 32 $BaTiO_3$ supercell obtained by the MACE + MACEField (blue) and the MACELES + MACEField model (green).

A closer examination of the switching pathway reveals that polarization reversal proceeds through a nearly homogeneous mechanism as shown in Figure 22, similar to the behavior observed at 270 K in Section III B. One possible explanation is that the MACELES model predicts a slightly smaller c/a ratio, as discussed in our previous study[43]. The reduced tetragonality may lower the energy barrier between the c- and a/b-polarized states, making transient reversed domains more susceptible to thermally activated annihilation before they can evolve into stable DW.

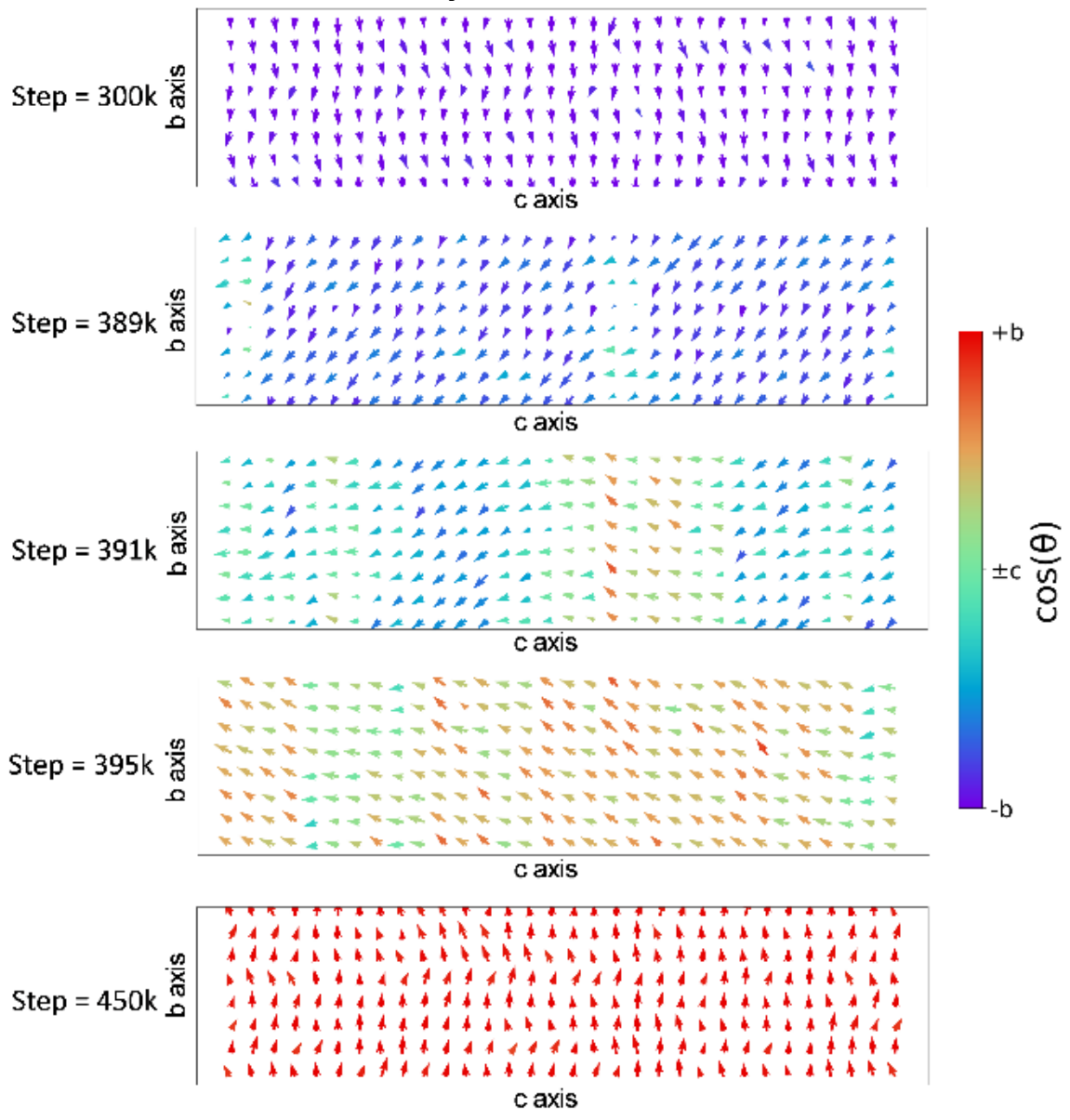

**Figure 22** Evolution of Ti displacement vectors in the 8 × 8 × 32 supercell at selected MD steps under a triangular electric field with a sweep rate of 1 kV/(cm*ps) applied along the b axis at 250 K by MACELES + MACEField model.

In contrast, when a higher electric-field sweep rate (2 kV/(cm*ps)) is applied, DW-mediated switching could be recovered, as shown in Figure 23. This result indicates that the qualitative switching mechanism remains unchanged regardless of whether long-range electrostatic interactions are explicitly included in the interatomic potential. Instead, the switching pathway appears to be primarily governed by the kinetic competition between DW nucleation/growth and thermally activated annihilation.

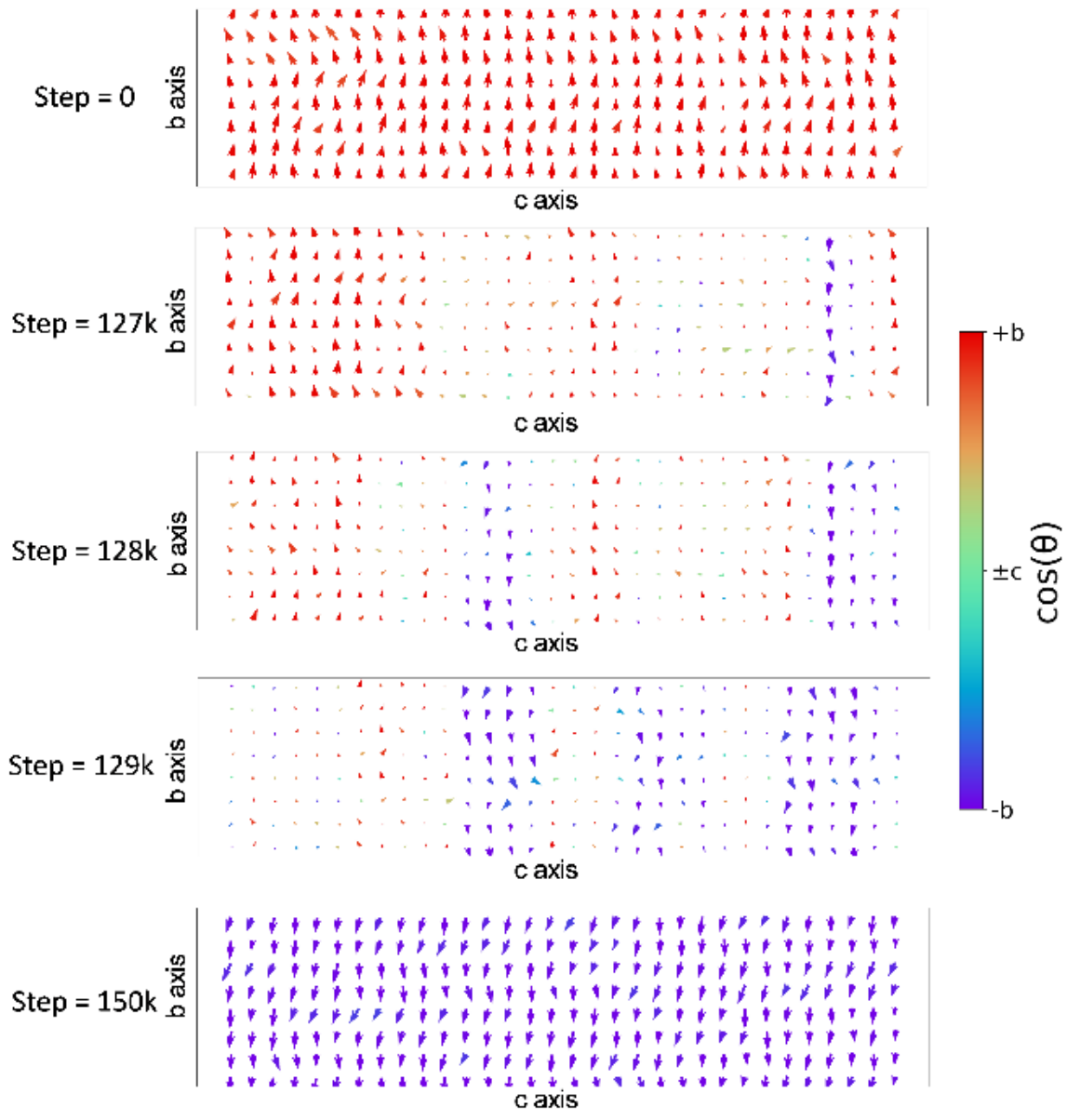

**Figure 23** Evolution of Ti displacement vectors in the 8 × 8 × 32 supercell at selected MD steps under a triangular electric field with a sweep rate of 2 kV/(cm*ps) applied along the b axis at 250 K by MACELES + MACEField model.

*Contact author: poyen@iis.u-tokyo.ac.jp
†Contact author: teru@iis.u-tokyo.ac.jp

[1] A. Jain, Y. Wang, L. Shi, Recent developments in BaTiO3 based lead-free materials for energy storage applications, JOURNAL OF ALLOYS AND COMPOUNDS, 928, 167066 (2022).
[2] M. Acosta, N. Novak, V. Rojas, S. Patel, R. Vaish, J. Koruza, G. Rossetti, J. Rödel, BaTiO3-based piezoelectrics: Fundamentals, current status, and perspectives, APPLIED PHYSICS REVIEWS, 4, 041305 (2017).
[3] Y. Jiang, Z. Tian, P. Kavle, H. Pan, L. Martin, Effect of fabrication processes on BaTiO3 capacitor properties, APL MATERIALS, 12, 041116 (2024).
[4] M. Richman, P. Rulis, A. Caruso, Ferroelectric system dynamics simulated by a second-order Landau model, JOURNAL OF APPLIED PHYSICS, 122, 094101 (2017).
[5] P. Kim, S. Han, Y. Kang, BaGe2P2 and BaGe2As2 pnictides as promising ferroelectric semiconductors for thin film solar cell, PHYSICAL CHEMISTRY CHEMICAL PHYSICS, 27, 17514-17520 (2025).
[6] Y. Li, F. Li, Domain switching criterion for ferroelectric single crystals under uni-axial electromechanical loading, MECHANICS OF MATERIALS, 93, 246-256 (2016).
[7] W. MERZ, THE ELECTRIC AND OPTICAL BEHAVIOR OF BATIO3 SINGLE-DOMAIN CRYSTALS, PHYSICAL REVIEW, 76, 1221-1225 (1949).
[8] L. Gigli, M. Veit, M. Kotiuga, G. Pizzi, N. Marzari, M. Ceriotti, Thermodynamics and dielectric response of BaTiO3 by data-driven modeling, NPJ COMPUTATIONAL MATERIALS, 8, 209 (2022).
[9] C. Cazorla, C. Escorihuela-Sayalero, J. Carrete, J. Iniguez-González, R. Rurali, Optical Control of the Thermal Conductivity in BaTiO3, ADVANCED FUNCTIONAL MATERIALS, 35, e2425424 (2025).
[10] H. Moriwake, C.A.J. Fisher, A. Kuwabara, T. Hashimoto, Anisotropic Permittivity of Tetragonal BaTiO3: A First-Principles Study, JAPANESE JOURNAL OF APPLIED PHYSICS, 50, 09NE02 (2011).
[11] P.-Y. Chen, T. Mizoguchi, Effect of uniaxial compressive stress on polarization switching and domain wall formation in tetragonal phase BaTiO3 via machine learning potential, Materials & Design, 265, 115851 (2026).
[12] H. Wu, J. Zhu, T. Zhang, Double hysteresis loops and large negative and positive electrocaloric effects in tetragonal ferroelectrics, PHYSICAL CHEMISTRY CHEMICAL PHYSICS, 17, 23897-23908 (2015).
[13] J. Lee, H. Song, D. Jeong, J. Ryu, Dielectric and ferroelectric property changes of BaTiO3 ceramics under in-plane stress application, JOURNAL OF ALLOYS AND COMPOUNDS, 1020, 179404 (2025).
[14] G. Deguchi, R. Kobayashi, H. Azuma, S. Ogata, M. Uranagase, S. Spreafico, Asymmetric Domain Nucleation from Dislocation Core in Barium Titanate: Molecular Dynamics Simulation Using Machine-Learning Potential through Active Learning, PHYSICA STATUS SOLIDI-RAPID RESEARCH LETTERS, 18, 2300292 (2024).
[15] J. Shieh, J. Yeh, Y. Shu, J. Yen, Hysteresis behaviors of barium titanate single crystals based on the operation of multiple 90° switching systems, MATERIALS SCIENCE AND ENGINEERING B-ADVANCED FUNCTIONAL SOLID-STATE MATERIALS, 161, 50-54 (2009).
[16] J. Liu, D. Yang, A. Suzana, S. Leake, I. Robinson, Electric field driven domain wall dynamics in BaTiO3 nanoparticles, PHYSICAL REVIEW B, 111, 054101 (2025).
[17] B. Jiang, Y. Bai, W. Chu, Y. Su, L. Qiao, Direct observation of two 90° steps of 180° domain switching in BaTiO3 single crystal under an antiparallel electric field, APPLIED PHYSICS LETTERS, 93, 152905 (2008).
[18] Z. Zhang, X. Qi, X. Duan, Two-step evolution mechanism of multi-domains in BaTiO3 single crystal investigated by in situ transmission electron microscopy, SCRIPTA MATERIALIA, 58, 441-444 (2008).
[19] X. Qi, H. Liu, X. Duan, In situ transmission electron microscopy study of electric-field-induced 90° domain switching in BaTiO3 single crystals, APPLIED PHYSICS LETTERS, 89, 092908 (2006).
[20] D. Lourens, M. Kwaaitaal, C. Davies, A. Kirilyuk, Time-resolved single-shot imaging of ultrafast polarization dynamics in barium titanate using pump-probe second-harmonic generation microscopy, REVIEW OF SCIENTIFIC INSTRUMENTS, 96, 073701 (2025).
[21] H. Azuma, T. Ogawa, S. Ogata, R. Kobayashi, M. Uranagase, T. Tsuzuki, F. Wendler, Unique temperature-dependence of polarization switching paths in ferroelectric BaTiO3: A molecular dynamics simulation study, ACTA MATERIALIA, 296, 121216 (2025).
[22] S. Falletta, A. Cepellotti, A. Johansson, C. Tan, M. Descoteaux, A. Musaelian, C. Owen, B. Kozinsky, Unified differentiable learning of electric response, NATURE COMMUNICATIONS, 16, 4031 (2025).
[23] Y. Shin, I. Grinberg, I. Chen, A. Rappe, Nucleation and growth mechanism of ferroelectric domain-wall motion, NATURE, 449, 881-U887 (2007).
[24] Y. Li, J. Wang, F. Li, Intrinsic polarization switching in BaTiO3 crystal under uniaxial electromechanical loading, PHYSICAL REVIEW B, 94, 184108 (2016).

*Contact author: poyen@iis.u-tokyo.ac.jp
†Contact author: teru@iis.u-tokyo.ac.jp

[25] P.-Y. Chen, T. Mizoguchi, Electric Field-Induced Phase Transitions and Hysteresis in Ferroelectric HfO2 Captured with Machine Learning Potential, MATERIALS TODAY ELECTRONICS, 17, 100228 (2026).
[26] P. Chen, K. Shibata, T. Mizoguchi, High precision machine learning force field development for BaTiO3 phase transitions, amorphous, and liquid structures, APL MACHINE LEARNING, 3, 036115 (2025).
[27] B.A. Martin, A.M. Ganose, V. Kapil, T. Li, K.T. Butler, General Learning of the Electric Response of Inorganic Materials, PRX INTELLIGENCE, 1, 013006 (2026).
[28] A. Kutana, K. Shimizu, S. Watanabe, R. Asahi, Representing Born effective charges with equivariant graph convolutional neural networks, SCIENTIFIC REPORTS, 15, 16719 (2025).
[29] K. Shimizu, R. Otsuka, M. Hara, E. Minamitani, S. Watanabe, Prediction of Born effective charges using neural network to study ion migration under electric fields: applications to crystalline and amorphous Li3PO4, SCIENCE AND TECHNOLOGY OF ADVANCED MATERIALS: METHODS, 3, 2253135 (2023).
[30] I. Batatia, P. Benner, Y. Chiang, A.M. Elena, D.P. Kovács, J. Riebesell, X.R. Advincula, M. Asta, M. Avaylon, W.J. Baldwin, F. Berger, N. Bernstein, A. Bhowmik, F. Bigi, S.M. Blau, V. Cărare, M. Ceriotti, S. Chong, J.P. Darby, S. De, F.D. Pia, V.L. Deringer, R. Elijošius, Z. El-Machachi, E. Fako, F. Falcioni, A.C. Ferrari, J.L.A. Gardner, M.J. Gawkowski, A. Genreith-Schriever, J. George, R.E.A. Goodall, J. Grandel, C.P. Grey, P. Grigorev, S. Han, W. Handley, H.H. Heenen, K. Hermansson, C.H. Ho, S. Hofmann, C. Holm, J. Jaafar, K.S. Jakob, H. Jung, V. Kapil, A.D. Kaplan, N. Karimitari, J.R. Kermode, P. Kourtis, N. Kroupa, J. Kullgren, M.C. Kuner, D. Kuryla, G. Liepuoniute, C. Lin, J.T. Margraf, I.-B. Magdău, A. Michaelides, J.H. Moore, A.A. Naik, S.P. Niblett, S.W. Norwood, N. O'Neill, C. Ortner, K.A. Persson, K. Reuter, A.S. Rosen, L.A.M. Rosset, L.L. Schaaf, C. Schran, B.X. Shi, E. Sivonxay, T.K. Stenczel, C. Sutton, V. Svahn, T.D. Swinburne, J. Tilly, C.v.d. Oord, S. Vargas, E. Varga-Umbrich, T. Vegge, M. Vondrák, Y. Wang, W.C. Witt, T. Wolf, F. Zills, G. Csányi, A foundation model for atomistic materials chemistry, The Journal of Physical Chemistry, 163, 184110 (2025).
[31] I. Batatia, D. Kovács, G. Simm, C. Ortner, G. Csányi, MACE: Higher Order Equivariant Message Passing Neural Networks for Fast and Accurate Force Fields, ADVANCES IN NEURAL INFORMATION PROCESSING SYSTEMS 35, NEURIPS 2022, 2022.
[32] X. Gonze, C. Lee, Dynamical matrices, born effective charges, dielectric permittivity tensors, and interatomic force constants from density-functional perturbation theory, PHYSICAL REVIEW B, 55, 10355-10368 (1997).
[33] P. Giannozzi, S. De Gironcoli, P. Pavone, S. Baroni, Ab initio calculation of phonon dispersions in semiconductors, Physical Review B, 43, 7231 (1991).
[34] G. Kresse, J. Furthmuller, Efficient iterative schemes for ab initio total-energy calculations using a plane-wave basis set, PHYSICAL REVIEW B, 54, 11169-11186 (1996).
[35] G. KRESSE, J. HAFNER, AB-INITIO MOLECULAR-DYNAMICS SIMULATION OF THE LIQUID-METAL AMORPHOUS-SEMICONDUCTOR TRANSITION IN GERMANIUM, PHYSICAL REVIEW B, 49, 14251-14269 (1994).
[36] G. Kresse, J. Furthmuller, Efficiency of ab-initio total energy calculations for metals and semiconductors using a plane-wave basis set, COMPUTATIONAL MATERIALS SCIENCE, 6, 15-50 (1996).
[37] A. Larsen, J. Mortensen, J. Blomqvist, I. Castelli, R. Christensen, M. Dulak, J. Friis, M. Groves, B. Hammer, C. Hargus, E. Hermes, P. Jennings, P. Jensen, J. Kermode, J. Kitchin, E. Kolsbjerg, J. Kubal, K. Kaasbjerg, S. Lysgaard, J. Maronsson, T. Maxson, T. Olsen, L. Pastewka, A. Peterson, C. Rostgaard, J. Schiotz, O. Schütt, M. Strange, K. Thygesen, T. Vegge, L. Vilhelmsen, M. Walter, Z. Zeng, K. Jacobsen, The atomic simulation environment-a Python library for working with atoms, JOURNAL OF PHYSICS-CONDENSED MATTER, 29, 273002 (2017).
[38] NVIDIA, cuEquivariance, version 0.10.0, 2026. [Online]. Available: https://github.com/NVIDIA/cuEquivariance. [Accessed: May 19, 2026].
[39] S. MELCHIONNA, G. CICCOTTI, B. HOLIAN, HOOVER NPT DYNAMICS FOR SYSTEMS VARYING IN SHAPE AND SIZE, MOLECULAR PHYSICS, 78, 533-544 (1993).
[40] S. Melchionna, Constrained systems and statistical distribution, PHYSICAL REVIEW E, 61, 6165-6170 (2000).
[41] C. SHANNON, A MATHEMATICAL THEORY OF COMMUNICATION, BELL SYSTEM TECHNICAL JOURNAL, 27, 379-423 (1948).
[42] D. Zhao, T. Lenz, G. Gelinck, P. Groen, D. Damjanovic, D. de Leeuw, I. Katsouras, Depolarization of multidomain ferroelectric materials, NATURE COMMUNICATIONS, 10, 2547 (2019).
[43] P.-Y. Chen, T. Mizoguchi, Long-range interaction effects on the phase transition, mechanical effect, and electric field response of BaTiO3 by machine learning potentials, APL MACHINE LEARNING, 4, 036101 (2026).

*Contact author: poyen@iis.u-tokyo.ac.jp
†Contact author: teru@iis.u-tokyo.ac.jp

[44] D. Kim, X. Wang, S. Vargas, P. Zhong, D. King, T. Inizan, B. Cheng, A Universal Augmentation Framework for Long-Range Electrostatics in Machine Learning Interatomic Potentials, JOURNAL OF CHEMICAL THEORY AND COMPUTATION, 21, 12709-12724 (2025).
[45] D. King, D. Kim, P. Zhong, B. Cheng, Machine learning of charges and long-range interactions from energies and forces, NATURE COMMUNICATIONS, 16, 8763 (2025).
[46] B. Cheng, Latent Ewald summation for machine learning of long-range interactions, NPJ COMPUTATIONAL MATERIALS, 11, 80 (2025).
[47] P. Zhong, D. Kim, D. King, B. Cheng, Machine learning interatomic potential can infer electrical response, NPJ COMPUTATIONAL MATERIALS, 11, 384 (2025).

*Contact author: poyen@iis.u-tokyo.ac.jp
†Contact author: teru@iis.u-tokyo.ac.jp